\documentclass[sigconf, nonacm]{acmart}
\usepackage{enumitem}
\usepackage{booktabs}
\usepackage{adjustbox}
\usepackage{multirow}
\usepackage{amsthm}
\usepackage{algorithm}
\usepackage{algpseudocode}

\newtheorem{definition}{Definition}[section] 
\newcommand{\vpara}[1]{\vspace{0.03in}\noindent \textbf{#1 }}

\newcommand\PHP{\textit{pg\_hint\_plan}}
\newcommand\Model{LLMOpt}
\newcommand\Baoarm{\textit{Bao\_arm}}

\newcommand\vldbdoi{XX.XX/XXX.XX}
\newcommand\vldbpages{XXX-XXX}
\newcommand\vldbvolume{14}
\newcommand\vldbissue{1}
\newcommand\vldbyear{2020}
\newcommand\vldbauthors{\authors}
\newcommand\vldbtitle{\shorttitle} 
\newcommand\vldbavailabilityurl{URL_TO_YOUR_ARTIFACTS}
\newcommand\vldbpagestyle{plain}

\usepackage{amsmath}
\begin{document}
\title{A Query Optimization Method Utilizing Large Language Models}

\author{
    Zhiming Yao$^{1,2}$,
    Haoyang Li$^{1,2}$,
    Jing Zhang$^{1,3\dagger}$,
    Cuiping Li$^{1,3}$,
    Hong Chen$^{1,3}$    
}
\affiliation{
    $^1$ School of Information, Renmin University of China, Beijing, China, \\
    $^2$ Key Laboratory of Data Engineering and Knowledge Engineering, MOE, China,\\
    $^3$ Engineering Research Center of Database and Business Intelligence, MOE, China
}
\thanks{$\dagger$ Jing Zhang is the corresponding author.}

\email{
  {yaojimmy2005, lihaoyang.cs, zhang-jing, licuiping, chong}@ruc.edu.cn
}



\begin{abstract}



Query optimization is a critical task in database systems, focused on determining the most efficient way to execute a query from an enormous set of possible strategies. Traditional approaches rely on heuristic search methods and cost predictions, but these often struggle with the complexity of the search space and inaccuracies in performance estimation, leading to suboptimal plan choices.
This paper presents \Model, a novel framework that leverages Large Language Models (LLMs) to address these challenges through two innovative components: (1) LLM for Plan Candidate Generation (\Model(G)), which eliminates heuristic search by utilizing the reasoning abilities of LLMs to directly generate high-quality query plans, and (2) LLM for Plan Candidate Selection (\Model(S)), a list-wise cost model that compares candidates globally to enhance selection accuracy. To adapt LLMs for query optimization, we propose fine-tuning pre-trained models using optimization data collected offline.
Experimental results on the JOB, JOB-EXT, and Stack benchmarks show that \Model(G) and \Model(S) outperform state-of-the-art methods, including PostgreSQL, BAO, and HybridQO. Notably, \Model(S) achieves the best practical performance, striking a balance between plan quality and inference efficiency.

\end{abstract}

\maketitle

\pagestyle{\vldbpagestyle}
\begingroup\small\noindent\raggedright\textbf{PVLDB Reference Format:}\\
\vldbauthors. \vldbtitle. PVLDB, \vldbvolume(\vldbissue): \vldbpages, \vldbyear.\\
\href{https://doi.org/\vldbdoi}{doi:\vldbdoi}
\endgroup
\begingroup
\renewcommand\thefootnote{}\footnote{\noindent
This work is licensed under the Creative Commons BY-NC-ND 4.0 International License. Visit \url{https://creativecommons.org/licenses/by-nc-nd/4.0/} to view a copy of this license. For any use beyond those covered by this license, obtain permission by emailing \href{mailto:info@vldb.org}{info@vldb.org}. Copyright is held by the owner/author(s). Publication rights licensed to the VLDB Endowment. \\
\raggedright Proceedings of the VLDB Endowment, Vol. \vldbvolume, No. \vldbissue\ %
ISSN 2150-8097. \\
\href{https://doi.org/\vldbdoi}{doi:\vldbdoi} \\
}\addtocounter{footnote}{-1}\endgroup

\ifdefempty{\vldbavailabilityurl}{}{
\vspace{.3cm}
\begingroup\small\noindent\raggedright\textbf{PVLDB Artifact Availability:}\\
The source code, data, and/or other artifacts have been made available at \url{https://github.com/lucifer12346/LLMOpt}.
\endgroup
}

\section{Introduction}
\label{section:intro}
\begin{figure}
    \centering
    \includegraphics[width=1\linewidth]{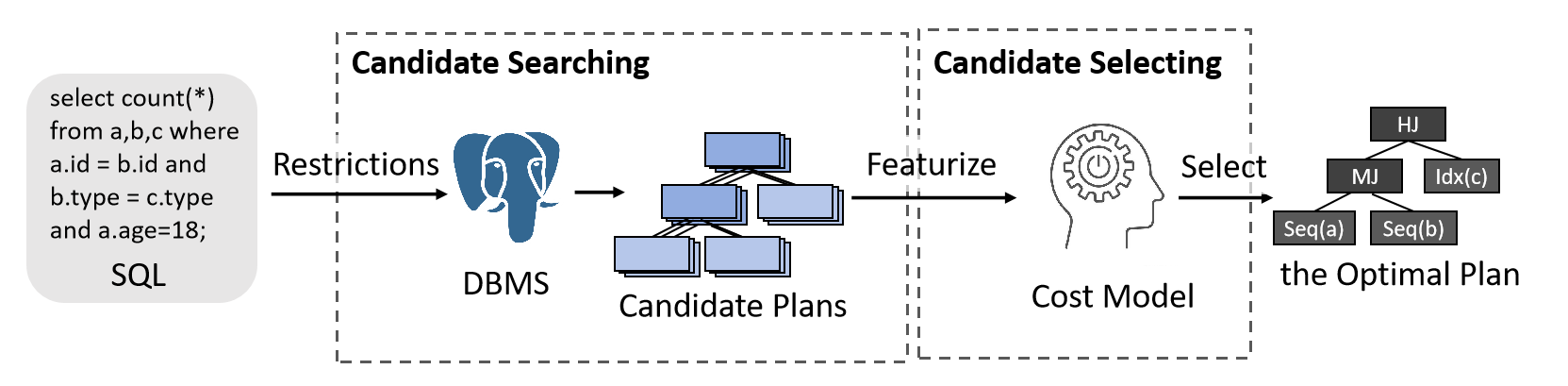}
    \caption{The Search-and-Select Pipeline Query Optimization}
    \label{fig:pipeline}
\end{figure}

In modern enterprises, databases execute vast numbers of queries daily to support critical business operations. Query optimization, a fundamental task in database management, aims to accelerate query execution by identifying efficient execution plans. These plans are selected from a search space of combinatorial complexity $O(n!)$, where $n$ represents the number of tables in the query. To navigate this challenge, database management systems (DBMS) such as the widely-used open-source PostgreSQL (PG) employ dynamic programming algorithms. These algorithms iteratively construct query plans by joining new tables to existing sub-plans (partial combinations of the query’s tables) to generate candidate plans. During this process, suboptimal candidates are pruned based on performance estimates, retaining only the most promising plans for further exploration.
The effectiveness of this approach hinges on two key challenges: rapidly exploring the search space and accurately estimating the performance of candidate plans. Prior research has proposed enhancements to query optimization through heuristic search strategies and learning-based performance estimation techniques \cite{marcus2021bao,chen2023leon,marcus2019neo,yang2022balsa,zhu2023lero}. These methods typically follow a search-and-select pipeline (Figure \ref{fig:pipeline}), where candidate plans are searched and the most promising ones are selected. However, each stage of this pipeline faces limitations that impact overall optimization efficacy.



\begin{figure}
    \centering
    \includegraphics[width=1\linewidth]{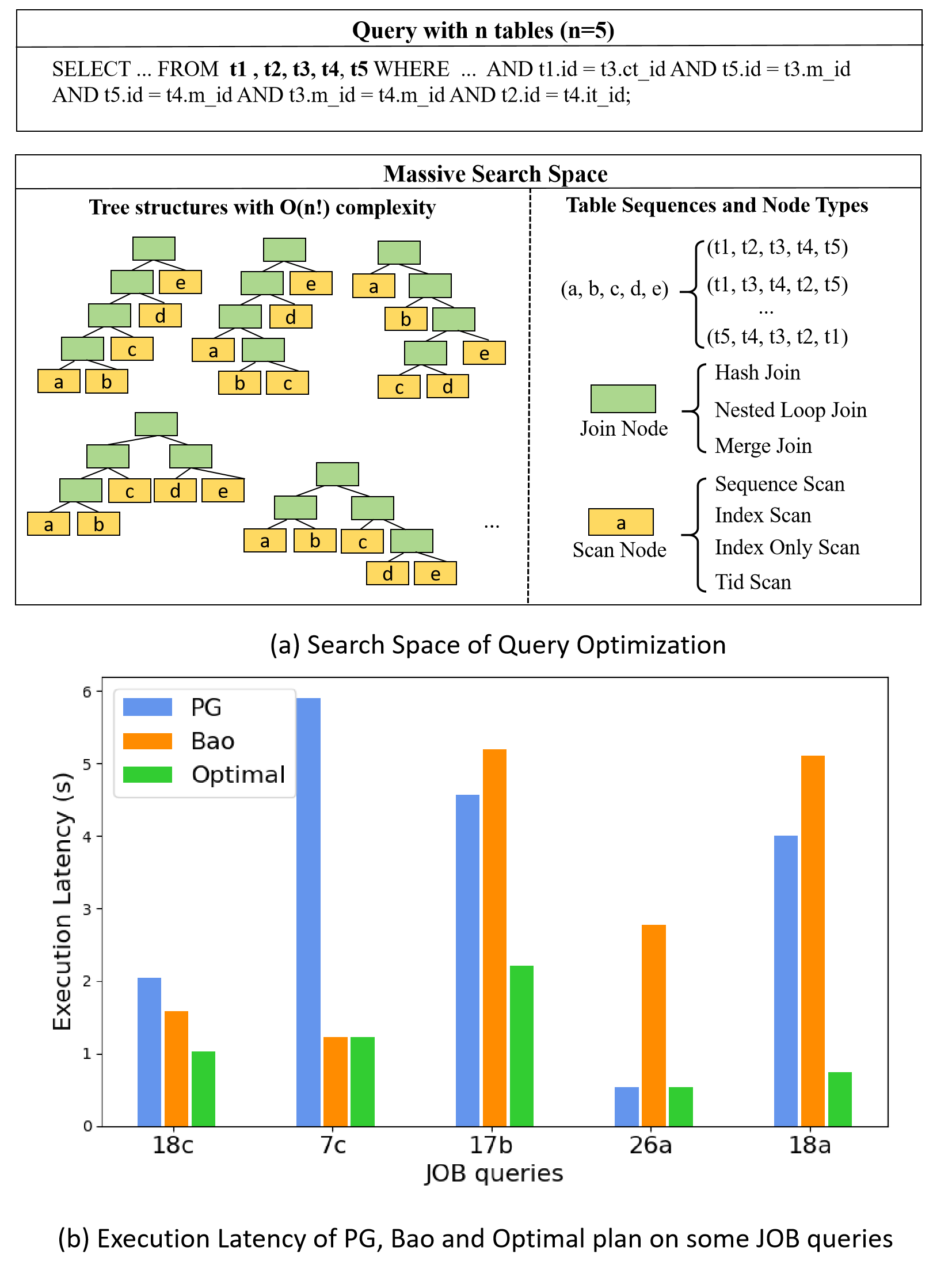}
    \caption{Limitations of Existing Methods — (a) Heuristic search strategies inefficiently navigate large query plan spaces, compounded by (b) suboptimal plan selection due to inaccurate cost models.}
    \label{fig:example}
\end{figure}

\vpara{(1) Massive Search Space.} A query corresponds to an enormous number of query plan candidates. Specifically, a query with \( n \) tables could derive \( O(n!) \) query plans. Figure~\ref{fig:example}(a) illustrates the possible tree structures and node values for a query. In the figure, yellow leaf nodes represent tables with various scan operations, such as sequential scan, index scan, index-only scan, and tid scan. Meanwhile, green non-leaf nodes denote join operations, including hash join, nested loop join, and merge join. The combination of tree structures, leaf nodes, and non-leaf nodes results in a massive number of plan tree candidates: 1) The number of distinct tree structures follows the Catalan numbers, which grow at a complexity of \( O(n!) \) \cite{stanley2015catalan}. For instance, as shown in Figure~\ref{fig:example}(a), a query with 5 tables corresponds to \( \text{Catalan}(4) = 14 \) unique tree structures. 2) Each tree structure can have multiple permutations of table sequences for its leaf nodes. 3) Furthermore, each leaf node has multiple scan operation options, and each non-leaf node has multiple join operation options.

Conventional methods~\cite{marcus2021bao, zhao2022queryformer} rely on heuristic search algorithms to explore candidate plans and use cost models to prune the search space. However, these approaches risk eliminating the optimal plan due to the limitations of heuristic search strategies in handling the combinatorial explosion and inaccuracies  inherent  in the cost model.

\vpara{(2) Poor Selection Performance.} Even when the set of searched query plan candidates contains the optimal plan, cost models such as the Tree-CNN used in BAO~\cite{marcus2021bao} or the Tree-LSTM used in HybridQO~\cite{yu2022cost} may fail to select the optimal plan from the candidates. Figure~\ref{fig:example}(b) compares the execution latency of PG, BAO, and the Optimal method on five queries from the join-order-benchmark (JOB). The Optimal method shares the same candidate plans as BAO, but while BAO relies on a learned cost model to estimate the latency of each plan, the Optimal method executes every candidate plan on the database to obtain the true execution latency. Thus, the Optimal method serves as an upper bound for BAO's performance.
As shown in the Figure ~\ref{fig:example}(b), for query 18c, even with the same set of candidate plans, BAO results in worse performance than the Optimal method, leaving a 20\% optimization gap. For queries 17b, 18a, and 26a, although the Optimal method outperforms the default PG, BAO's selected query plan performs worse than the plan derived by PG, indicating that BAO's cost model struggles to select the optimal plan. 

To address the challenges discussed above, large language models (LLMs) have emerged as a promising solution, thanks to their powerful natural language understanding and generation capabilities. 
On one hand, LLMs are widely used to generate textual content for various natural language processing (NLP) tasks. Given that natural language involves an even larger search space than query plans, it is reasonable to conjecture that LLMs can effectively generate optimal plans from the massive query plan search space. 
On the other hand, LLMs have demonstrated exceptional understanding and reasoning abilities~\cite{hurst2024gpt,guo2025deepseek}, particularly in tasks involving coding~\cite{chew2023llm,jiang2024survey} and mathematics~\cite{imani2023mathprompter}. Similarly, query optimization requires strong reasoning capabilities to deduce the optimal query plan based on the given query and database state. 
Moreover, LLMs have already proven useful in various database-related tasks. For instance, researchers have explored their application in text-to-SQL~\cite{dong2023c3,li2024codes}, knob tuning~\cite{huang2024llmtune,giannakouris2024demonstrating}, and query rewriting~\cite{ma2023query,liu2024query}. Drawing inspiration from LLMs' natural language understanding and generation capabilities, as well as their success in database tasks, we propose leveraging LLMs for query optimization.

\vpara{Solutions:} We propose \Model, a \textbf{L}arge \textbf{L}anguage \textbf{M}odel \textbf{Opt}imizer, which explores two potential approaches for utilizing LLMs as query optimizers:

\textbf{(1) LLM for Plan Candidate Generation.}
The first approach, referred to as \Model(G), replaces the candidate searching component in the query optimization pipeline (as illustrated in Figure~\ref{fig:pipeline}) with an LLM. Instead of relying on heuristic search algorithms, the LLM directly generates query plan candidates. The LLM is designed to perform multiple generations to produce several candidate plans. These candidates are then evaluated using the conventional method~\cite{marcus2021bao}, which utilizes a Tree-CNN-based cost model to select the optimal query plan. Unlike heuristic search algorithms that explicitly traverse or prune the state space, LLMs inherently narrow the plan candidate space through parameterized probability distributions, offering an implicit and effective search mechanism.

\textbf{(2) LLM for Plan Candidate Selection.}
The second approach, referred to as \Model(S),  replaces the cost model used for candidate selection in the pipeline with an LLM-based list-wise cost model. This list-wise model accepts multiple candidate plans simultaneously as input and selects the optimal plan from them. Unlike traditional approaches that rely on point-wise estimation (estimating each candidate independently)~\cite{marcus2019neo,yang2022balsa,marcus2021bao,yu2022cost} or pairwise estimation (comparing candidates locally)~\cite{chen2023leon,zhu2023lero}, the list-wise model conducts a global comparison across all candidates, leading to more informed and effective decisions.

\textbf{Enabling LLMs for Query Optimization.} Achieving these two approaches with existing pre-trained LLMs (PLMs) is challenging because standard PLMs struggle to fully comprehend query plans and cannot inherently determine what constitutes a good query plan. Determining the quality of a query plan requires interaction with the database, which cannot be learned from the text corpora typically used to train PLMs. To address these limitations, we propose a method to collect query optimization data offline by interacting with the database.
Using the collected data, we fine-tune a PLM into two distinct models: $LLM^{G}$ and $LLM^{S}$. These models serve as a plan generator and a plan selector within the query optimization pipeline, forming the proposed \Model(G) and \Model(S), respectively.

The primary contributions  can be summarized as follows:

\begin{itemize}[leftmargin=1em]
\item \textbf{LLM-Driven Query Optimization Model:} We propose \Model, which leverages LLMs for query optimization. In the search-and-selection query optimization pipeline, we replace the search component with an LLM-based candidate plan generator that narrows the candidate space efficiently using parameterized probability distributions. We also replace the selection component with an LLM-based list-wise cost model that performs a global comparison of all candidate plans for more effective plan selection.

\item \textbf{Extensive Benchmark Evaluation:} We evaluate \Model\ on three benchmarks: JOB, JOB-EXT, and Stack, comparing our models to state-of-the-art baselines like PG default, BAO, and HybridQO. Both \Model(G) and \Model(S) outperform all baselines, including tuning-free PLMs like GPT-4o, demonstrating the critical importance of fine-tuning PLMs with the collected data.

\item \textbf{Performance Insight:} $LLM^{G}$ in \Model(G) generates plans with lower mean and variance in execution latency compared to heuristic search methods, showing its potential as a search component. $LLM^{S}$ in \Model(S) outperforms smaller models like Tree-CNN and Tree-LSTM for plan selection. However, combining them does not always outperform pure \Model(G) or \Model(S) on certain benchmarks, such as JOB. This result suggests that while $LLM^{G}$ can generate more optimal plans, $LLM^{S}$ may insufficiently select them due to a mismatch between the distribution of $LLM^{G}$'s generated plans and $LLM^{S}$'s learned distribution. This highlights the need for further improvements in $LLM^{S}$ to address this gap in the future. Finally, from the current experimental results, we find that pure \Model(S) is the best practical choice due to its ability to produce high-performing query plans while maintaining efficient end-to-end latency, even with the additional LLM selection cost.
\end{itemize}

\section{Related Work}
\subsection{Learning-based Query Optimization}
While conventional query optimizers still have significant room for improvement, machine learning models has increasingly been applied to query optimization in recent years. Beginning with Neo~\cite{marcus2019neo} and BAO~\cite{marcus2021bao}, researchers use machine learning models to either replace traditional optimizer in DBMS or enhance its performance for more efficient query execution.

\subsubsection{Replacing Traditional Optimizers}

The traditional optimizer in PG searches query plan by iteratively traversing possible sub-plans.
Relying on expert-defined heuristic rules, it estimates cost of each sub-plan and prunes those with high cost to reduce searching space. This process continues iteratively until a complete plan with all tables connected is found.  


Neo~\cite{marcus2019neo} and Balsa~\cite{yang2022balsa} replace traditional optimizers by using machine learning models instead of heuristic cost estimators, preventing the pruning of potentially optimal sub-plans. Both use Tree-CNN~\cite{mou2016convolutional} to predict execution latency from sub-plans. Neo relies on large amounts of real-world execution data for training, while Balsa starts with a conventional cost model and refines its predictions by learning from real-world optimizers like PG. Overall, these methods replace traditional optimizers and discard the conventional cost model in the search for optimal plans.

\subsubsection{Improving Traditional Optimizers}

Methods such as BAO~\cite{marcus2021bao} and HybridQO~\cite{yu2022cost}, on the other hand, partially rely on expert-defined rules in traditional optimizers while leveraging machine learning models to enhance performance. These methods recognize the capability of traditional optimizers to identify optimal plans but refine their approach to more effectively capitalize on this potential. Specifically, these methods constrain the searching process of traditional optimizers to obtain multiple plan candidates and employ a cost model to select the optimal one from plan candidates.

BAO~\cite{marcus2021bao}, for instance, restricts the optimizer from using certain type of scans or joins by adjusting database knobs, treating each group of knobs as an arm in a contextual multi-armed bandit problem. A Tree-CNN is then utilized to predict which particular plan is likely to yield the best performance. QueryFormer~\cite{zhao2022queryformer}, similar to BAO, adopts tree-structured transformer to represent query plans and estimate the execution latency of each plan candidate.
HybridQO~\cite{yu2022cost}, as a hybrid optimizer, employs a cost-based Monte Carlo Tree Search (MCTS) algorithm to guide the join order of the traditional optimizer and leverage a learning-based model Tree-LSTM model to select the optimal plan.

In conclusion, learning-based optimizers either tries to entirely replace traditional optimizers or enhance their performance with machine learning models.
However, a recent study\cite{lehmann2023your} highlights that despite leveraging machine learning models, existing methods still face significant end-to-end overhead, as execution latency remains excessively high, compounded by the additional cost of model inference.

\subsection{Large Language Models}
Large Language Models (LLMs) are powerful machine learning models based on the transformers structure~\cite{vaswani2017attention}. As the depth and width of transformers layers increase, the number of parameters in LLMs scales to billions and trillions. In recent years, LLMs such as GPT-4~\cite{openai2023gpt4} and DeepSeek-v3~\cite{liu2024deepseek} have demonstrated remarkable capability in natural language understanding (NLU) and natural language generation (NLG). 
Meanwhile, open-source LLMs such as LlaMA~\cite{dubey2024llama} and Mistral~\cite{jiang2023mistral} enable researchers to further fine-tune these models for specific tasks. 

Pre-trained language models (PLMs) exhibit strong generalization capabilities across various natural language processing (NLP) tasks, including code generation and question answering ~\cite{naveed2023comprehensive,xu2022systematic,daull2023complex}. However, in downstream applications, LLMs struggle to strictly follow user instructions and generate invalid or inaccurate outputs~\cite{ouyang2022training}.
Fine-tuning is an effective strategy for adapting LLMs to specific tasks, ensuring that LLMs are able to follow instructs and generate output in desired format.


\subsection{LLM for Database Tasks}
Given the strong capability in NLU and NLG, LLMs have been widely used in various database-related tasks.
For the text-to-sql task, numerous studies~\cite{pourreza2024din,dong2023c3,li2024codes, gao2023text} have demonstrated the impressive performance of LLMs~\cite{shi2024survey}. DIN-SQL~\cite{pourreza2024din} prompts GPT-4 with few-shot examples while CodeS~\cite{li2024codes} extends the training of StarCoder~\cite{li2023starcoder} and further fine-tunes it on augmented question-SQL pairs. For database knob tuning task, GPTuner~\cite{lao2023gptuner} prompts GPT-4 to reduce searching space and LLMTune~\cite{huang2024llmtune} fine-tunes Mistral-7B~\cite{jiang2023mistral} to provide a better starting point for base knob tuning optimizers. Beyond these applications, query rewriting task~\cite{ma2023query, liu2024query} utilizes LLMs and achieves a better performance.

\section{Preliminaries}

This section provides an overview of the conventional query optimization pipeline, which consists of the candidate searching and candidate selecting stages (Section~\ref{subsection:pipeline}). Since BAO's candidate searching method plays a critical role in our \Model, we describe it in Section~\ref{subsection:bao}. Additionally, we introduce the plugin tool \PHP, which is used in \Model\ to apply the optimized query plan to the database~ (Section~\ref{subsection:pg_hint_plan}). Finally, as \Model\ involves fine-tuning a PLM, we detail the key technical aspects of fine-tuning in Section~\ref{subsection:fine-tuning-LLMs}.


\subsection{Pipeline of Query Optimization}
\label{subsection:pipeline}

Existing methods primarily follow the pipeline shown in Figure ~\ref{fig:pipeline}, which consists of candidate searching and candidate selection. 

\vpara{Candidate Searching.} 
In the candidate searching stage, existing methods aim to obtain a diverse plan candidate set from which an optimal plan can be selected. 
BAO~\cite{marcus2021bao} and HybridQO~\cite{yu2022cost} influence the default query plan searching process of the conventional optimizer in a DBMS by imposing different restrictions. BAO adjusts database knobs to enable or disable specific types of scans or join operations, resulting in multiple distinct plans forming a plan candidate set. HybridQO, on the other hand, uses MCTS to find optimal or near-optimal table join orders and then employs \PHP\ to enforce these join orders on the conventional optimizer.


\vpara{Candidate Selecting.}
In the candidate selection stage, methods aim to pick the best plan from candidates identified earlier.
Conventional methods use expert-defined rules and database statistics to estimate cardinality and compute costs.
Learning-based methods leverage neural networks  to predict execution costs more accurately. These methods encode query details such as numerical properties and how tables and filters relate to each other. ~\cite{yang2022balsa,chen2023leon,marcus2019neo,yu2022cost}. They also take into account cardinality estimates and costs calculated by the DBMS.~\cite{marcus2021bao,zhao2022queryformer,marcus2019neo}.
To decide between plans, learning-to-rank loss is often used to compare and select the better option from two candidates.~\cite{chen2023leon,zhu2023lero}.

\subsection{Candidate Searching in BAO}
\label{subsection:bao}
BAO~\cite{marcus2021bao} is a well-performed method utilizing a Tree-CNN as model framework to select the best plan from plan candidates searched by the DBMS. As previously mentioned, BAO refines existing optimizers by adjusting DBMS knobs. In this paper, we refer to each unique combination of these knobs as a \Baoarm, since BAO treat each configuration as an arm in a multi-armed bandit problem.

In total, there are 48 distinct \Baoarm s, derived from six binary knobs that control whether the conventional optimizer can use specific types of scan or join methods during query plan searching. Specifically, each \Baoarm\ is represented as a six-dimensional binary vector, where a value of 0 disables the conventional optimizer from using a particular scan or join type throughout the search process.
By applying different \Baoarm s, DBMS can search diverse plan candidates, facilitating a more comprehensive search space for query optimization.


\subsection{Pg\_Hint\_Plan}
\label{subsection:pg_hint_plan}

We introduce \PHP~\cite{pghintplan}, an efficient plugin tool used in our method to apply the generated query plan to the PG DBMS for execution, which has been widely adopted in previous works~\cite{chen2023leon,yu2022cost,zhu2023lero}.
In the following sections, we briefly introduce \PHP\ and detail the revisions we made to enhance its effectiveness.


\vpara{How does \PHP\ work?}
\PHP\ allows users to customize the PG optimizer by imposing specific restrictions on the search for an optimal query plan. Specifically, \PHP\ enables users to control three key aspects of query optimization: the scan types for base tables (as base relations), the join methods for relations, and the join order of relations. These controls effectively define a complete query plan tree using three types of hints.
A set of hints in \PHP\ is represented as $h = \{S, J, L\}$, where $S$ consists of scan hints, $J$ consists of join hints, and $L$ is a leading hint. These components respectively restrict scan types, join methods, and join orders. For example, as illustrated in Figure~\ref{fig:PHP}, the hint \textit{SeqScan}(k) enforces the scan type for table k to be a sequential scan, while the hint \textit{NestedLoop}(k mk) specifies that the join type between tables k and mk must be a nested loop join. Additionally, the \textit{Leading} hint restricts the join order, ensuring that the specified tables are joined in the given sequence.
By combining these hints and injecting them into the DBMS by \PHP, users can precisely control the query plan that will be executed by the DBMS.

\vpara{How do we refine \PHP?}
\PHP\ accepts hints as input, injecting them into the DBMS to control the query plan. Ideally, the generated query plan should conform to the provided hints. However, the existing implementation of \PHP\ does not fully satisfy this requirement. Therefore, we refine \PHP\ to resolve this issue. We first analyze why \PHP\ fails to strictly follow the provided hints, resulting in an incorrect query plan. We discovered that when applying the sequential scan hint in \PHP, it invalidates indexes on the corresponding table throughout the entire search process. This forced invalidation can inadvertently alter join types, causing the query plan to deviate from the intended structure. To resolve this issue, we refine the implementation of the sequential scan hint to prevent the optimizer from searching for index scans, ensuring that join types remain unaffected.

In addition to solving the above issue, we also revise \PHP\ to improve its efficiency. The original \PHP\ does not effectively reduce the search space of the conventional optimizer. Even when a complete hint set is provided—fully specifying scan types, join types, and join order ($S, J, L$)—the optimizer still spends unnecessary time searching for an optimal plan. To eliminate this inefficiency, we refine \PHP\ so that when a complete hint set is provided, the optimizer bypasses the redundant search space and directly adheres to the specified hints.
The revised \PHP\ has been open-sourced for the community\footnote{\url{https://github.com/lucifer12346/pg_hint_plan_lucifer}}.

\subsection{Large Language Models Fine-tuning}
\label{subsection:fine-tuning-LLMs}
The fundamental architecture of modern LLMs, such as GPT series ~\cite{openai2023gpt4,lagler2013gpt2,floridi2020gpt} and LlaMA series~\cite{dubey2024llama,touvron2023llama}, is based on the transformer~\cite{vaswani2017attention}. The transformer utilizes multiple self-attention layers in both encoding and decoding components, allowing it to process input sequences and generate corresponding text outputs. Notably, modern LLMs are designed as decoder-only models, meaning that both training and inference rely on next-token-prediction. In this paradigm, each generated token is conditioned on the previously generated tokens as well as the input sequence:


\begin{equation}
    p_\theta(Y|x) = p_\theta(y_1,... y_m|X) = \prod_{t=1}^{m} p_\theta(y_t|X,y_{<t})
\end{equation}

\noindent where $p(\cdot)$ denotes the model's generation probability, $\theta$ denotes the model's learned parameters, $Y=\{y_1,...y_m\}$ is the output sequence, $y_t$ is the $t$-th token in the output sequence and $X$ denotes the input sequence. 

With next-token-prediction, modern LLMs minimize cross-entropy loss to enhance its ability of generating accurate outputs with maximum likelihood of the correct token:
\begin{equation}
\label{CEloss}
    Loss(\theta) = -\sum_{t=1}^{m}\log p_\theta(y_t|X,y_{<t})
\end{equation}

Following the loss function in Eq. ~\ref{CEloss}, a large-scale dataset $\mathbb{D} =\{(X,Y)\}$ is utilized for fine-tuning LLM. After fine-tuning, the LLM adapts to the specific downstream task corresponding to the dataset $\mathbb{D}$, enabling it to generate more accurate and task-specific outputs.


\section{Overview}
This section first defines the problem of query optimization and then presents the overview of the proposed \Model\ framework.
\subsection{Problem Definition}



\begin{definition}
\textit{\textbf{Query Optimization}.}
Given a query $q$ involving $n$ tables $T=\{t_1, t_2, ..., t_n\}$, query optimization is the process searching through a vast space of potential query plans to identify the optimal one. The complexity of this search arises from multiple factors: plan tree structures, table orderings and physical operations as illustrated in  Figure~\ref{fig:example} (a) in Section~\ref{section:intro}. Each query plan $p$ has an execution latency $l$, which is only obtainable by executing the query in a real-world DBMS. Typically, the goal of query optimization is to find the optimal plan $p^*$ that achieves the minimum execution latency $l^*$. To approximate execution latency without executing queries, existing methods often rely on a cost model to predict a cost $c$ for each plan, which guides the optimizer in searching the best plan.
\end{definition}

\begin{definition}

\textit{\textbf{Query Optimization utilizing LLMs.}}
For a given query $q$ and a database instance $d$, \Model\ replaces the candidate searching and candidate selection components in the traditional query optimization pipeline with two LLMs:
\begin{align}
    LLM^{G}_{\theta_G} (q, stats(d,q)) &\rightarrow h \\
    LLM^{S}_{\theta_S} (q, stats(d,q), H) &\rightarrow idx^*
\end{align}

\noindent where $LLM^G$ is a generative LLM responsible for generating query plan hints, $LLM^S$ is a selective LLM that selects the optimal query plan from plan candidates. The parameters of these models are represented by $\theta_G$ and $\theta_S$, respectively. The term $stats(\cdot)$ refers to database statistics relevant to $q$, $h$ is a sequence of hints associated with a specific query plan, and $H$ represents the set of hint candidates (i.e., sequences of hints) associated with the set of plan candidates $P$. Finally, $idx^*$ denotes the index of the optimal hint candidate $h^*$ within $H$. 

\end{definition}

\Model(G) refers to the query optimization method that uses $LLM^G$ to generate candidates, while \Model(S) refers to the method that uses $LLM^S$ to select the optimal plan.


\subsection{Framework of \Model}
\label{subsection:framework}
\begin{figure*}
    \centering
    \includegraphics[width=1\linewidth]{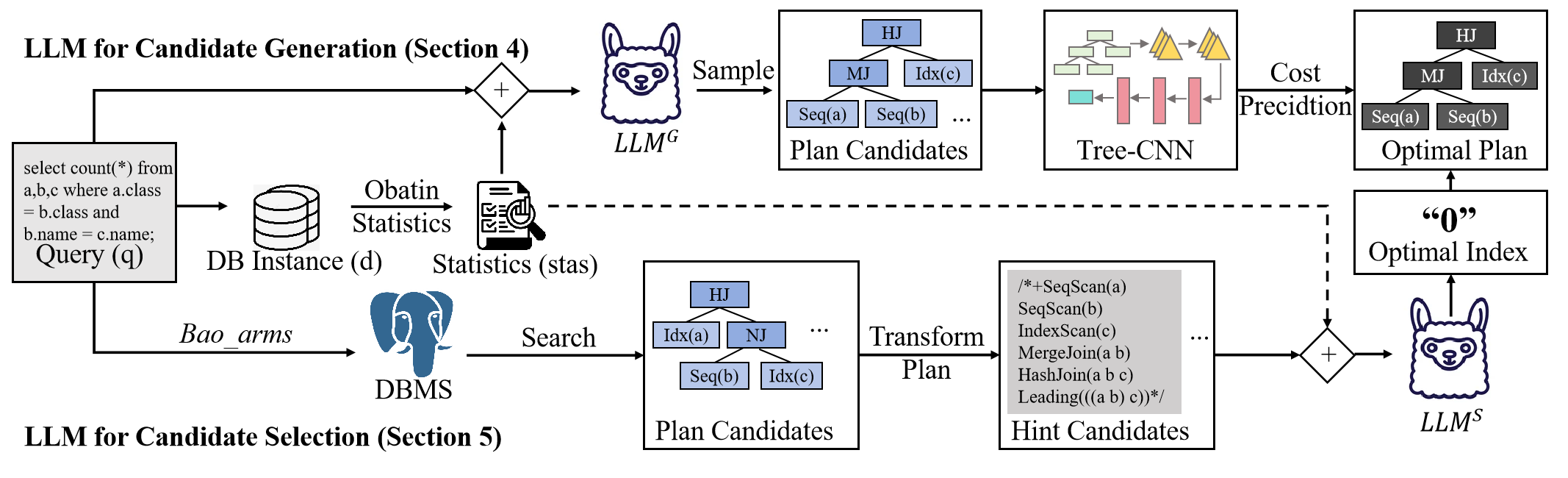}
    \caption{Framework of \Model, utilizing LLMs for candidate generation and candidate selection. }
    \label{fig:LLMOpt}
\end{figure*}

We propose \Model, a framework for leveraging \textbf{LLMs} in Query \textbf{Opt}imization, as illustrated in Figure~\ref{fig:LLMOpt}, which introduces two distinct approaches. We fine-tune a PLM into two distinct models, $LLM^{G}$ and $LLM^{S}$, which are used for plan candidate generation and selection in the query optimization pipeline, respectively.


\vpara{LLM for Plan Candidate Generation.}
To leverage the generative capabilities of LLMs in query optimization, our first approach, \Model(G), employs $LLM^G$ as a plan candidate generator. Given a query $q$, we construct an input sequence comprising $q$ along with the database statistics $stats$ relevant to $q$, which assists in generating the query plan. Since LLMs produce sequential text as output, we treat the hints corresponding to query plans, as detailed in Section~\ref{subsection:pg_hint_plan}, as the outputs of \Model(G). Replacing the traditional candidate searching process, we use the sampling mechanism of LLMs to generate multiple hints as plan candidates. Similar to BAO, we then employ a Tree-CNN to predict the execution latency of each plan candidate. Finally, the plan with the lowest predicted latency is selected as the optimal plan in this approach.

\vpara{LLM for Plan Candidate Selection.}
To address the limitations of learning-based methods in achieving high selection accuracy, we propose \Model(S), an approach that leverages $LLM^S$ as a powerful list-wise cost model in the query optimization process. Given a query $q$, we adopt BAO's method to search for plan candidates by applying \Baoarm s to the DBMS. Since LLMs require sequential inputs, we transform the tree-structured plans into sequential hints. We combine the query $q$, database statistics $stats$, and the hint candidates $H$ corresponding to the plan candidates as input to the LLM. The LLM is trained to identify the index of the optimal hint from the set of hint candidates. In this approach, the LLM generates a single token for each inference and performs inference only once per query, significantly reducing inference costs and accelerating the overall optimization process.

\section{LLM for Plan Candidate Generation}
\label{section:generate_model}
\Model(G) utilizes $LLM^G$ to generate plan candidates, as described in Algorithm ~\ref{alg:LLMQO_generate}. First, relevant statistics $stats$ are obtained (line 1). Since $LLM^G$ serves as a full replacement for conventional DBMS optimizers, it requires similar information to that used by traditional optimizers. These shared statistics, applicable to both \Model(G) and \Model(S), are detailed in Section ~\ref{subsection:statistics}. Next, $LLM^{G}$ generates candidates by taking a query and the statistics as inputs and producing multiple hints corresponding to plan candidates using a sampling strategy (line 2). Each hint candidate is then applied to the DBMS using \PHP\ (line 5), and Tree-CNN is used to predict the latency of each query plan (line 6). Finally, the hint $h^*$ corresponding to the plan with the lowest predicted latency is selected as the optimal plan (lines 7 to 10).



\begin{algorithm}[t]
\caption{\Model(G)}
\label{alg:LLMQO_generate}
\begin{algorithmic}[1] 
\Require query $q$, database instance $d$
\Ensure optimal hint $h^*$
\State $S$ $\gets$ $Obtain\_Statistics(q,d)$
\State hint candidates $H$ $\gets$ $LLM^G(q,stats)$
\State $c^* \gets$ a maximum value
\For{hint candiate $h$ in $H$}
\State plan $p$ $\gets$ DBMS$(q, h)$ 
\State predicted latency $c$ $\gets$ Tree-CNN($p$)
\If{$c<c^*$}
\State $c^* \gets c$
\State $h^* \gets h$
\EndIf
\EndFor
\State\Return $h^*$
\end{algorithmic}
\end{algorithm}



\vpara{LLM Inference Strategy.}
To enhance the diversity of generated hints, we adopt a sampling-based inference strategy. Inspired by inference scaling~\cite{xu2018scaling}, we hypothesize that sampling from $LLM^G$ is able to find better hints compared to greedy decoding. Specifically, we sample the output of the LLM with a relatively high temperature (t=1) for 16 times, introducing greater randomness and diversity in hint candidates $H$. This sampling strategy increases the likelihood of generating the optimal hint in the set of hint candidates. 


\vpara{Tree-CNN Selection.}
After generating multiple candidates, we employ the Tree-CNN model from BAO~\cite{marcus2021bao} to select the best one. For each hint $h$, we apply it to DBMS to obtain query plan $p$ and the Tree-CNN retains only the essential information—the types of join and scan nodes, the cardinality, and the cost at each node—while discarding other less relevant details such as memory allocation. Tree-CNN then predicts the latency for each query plan corresponding to each $h$ and selects the hint with the minimum prediction as the optimal one.


\section{LLM for Plan Candidate Selection}
\label{section:cost_model}

\begin{algorithm}[t]
    \caption{\Model(S)}
    \label{alg:LLM_select}
    \begin{algorithmic}[1]
        \Require query q, database instance d
        \Ensure optimal hint $h^*$
        
\State $stats$ $\gets$ $Obtain\_Statistics(q,d)$
\State hint candidates $H \gets []$
\For{\Baoarm\ in \Baoarm s}
    \State plan $p \gets \text{DBMS}(q, Bao\_arm)$
    \State hint $h \gets \text{Transform\_Plan}(p)$
    \State   Append $h$ to $H$
\EndFor
\State $idx^*$ $\gets$ $LLM^S(q,stats,H)$
\State $h^*$ $\gets$ $H[idx^*]$
\State \Return $h^*$

    \end{algorithmic}
\end{algorithm}

In this section, we introduce \Model(S), an approach that leverages $LLM^S$ for plan candidate selection, as outlined in Algorithm ~\ref{alg:LLM_select}. Following the methodology of BAO (described in Section ~\ref{subsection:bao}), we first search for a variety of candidate plans and transform these plans into sequential hints (lines 3 to 7). The transformation process is explained in Section ~\ref{subsection:label_collection}. Subsequently, we use $LLM^S$, a list-wise cost model, to select the optimal hint by generating the corresponding index (lines 8 and 9).


The core idea of \Model(S) is to use $LLM^S$ as a list-wise selection model, requiring only a single inference per query, which significantly enhances efficiency. As a list-wise cost model, $LLM^S$ has the advantage of producing a more accurate global ranking by considering interactions among all plans in the list, rather than focusing solely on individual or pairwise relationships. This allows $LLM^S$ to effectively capture the overall ranking structure.

Similar to $LLM^G$ described in Section ~\ref{section:generate_model}, the input sequence for $LLM^S$ consists of $(q, stats, H)$, where $H = \{h_1, ..., h_n\}$ represents the set of hint candidates corresponding to the plan candidates. The output of $LLM^S$ is a single token $idx^*$, which denotes the index of the predicted optimal hint $h^*$ in $H$. 

\section{Data Collection}
\label{section:data_collection}
\begin{figure}
    \centering
    \includegraphics[width=1\linewidth]{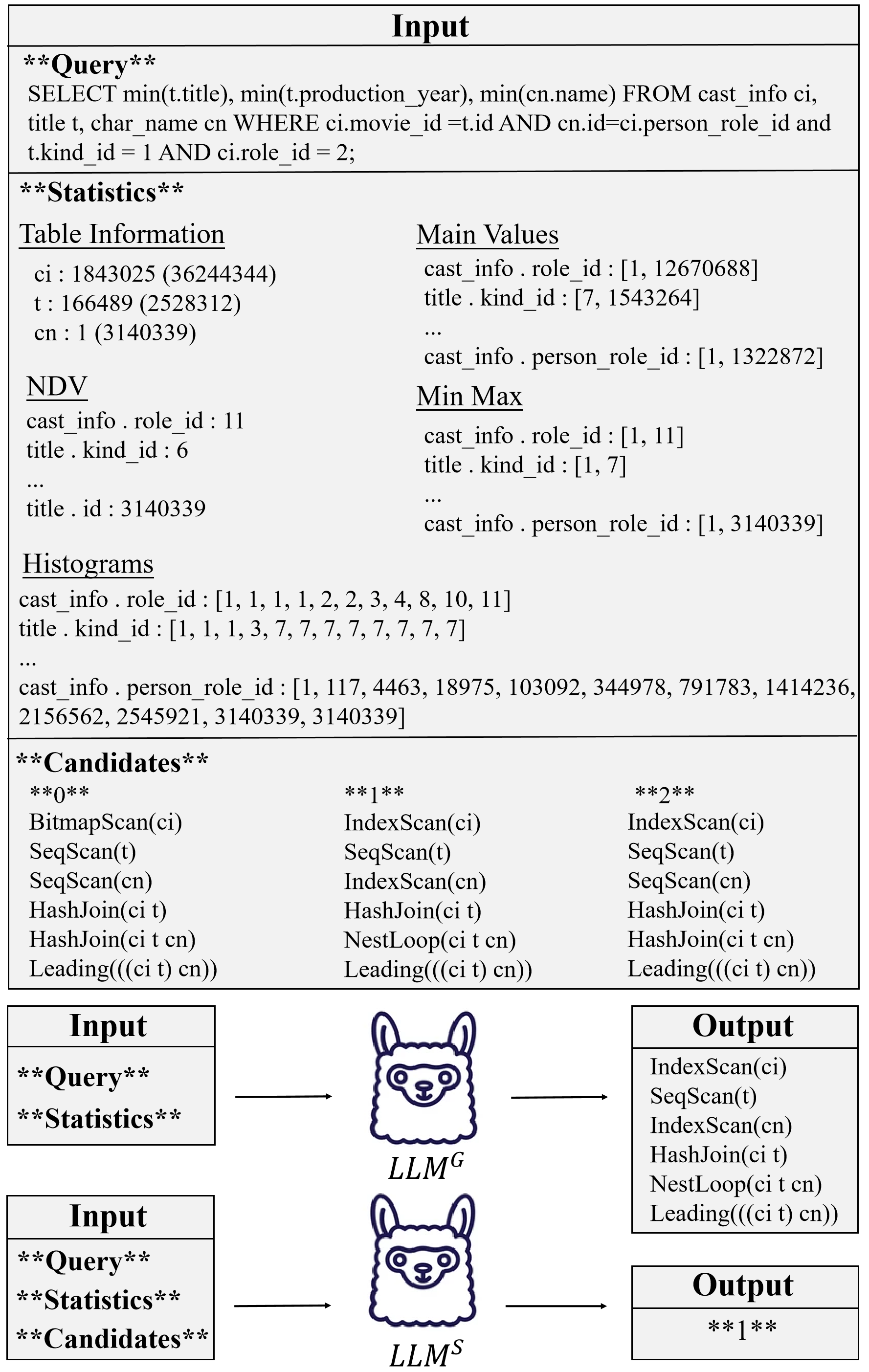}
    \caption{An Example of LLM Input and Output.}
    \label{fig:input_output}
\end{figure}

In this section, we outline the data collection process for \Model. Specifically, we gather the statistics $stats$, which are used as input for both $LLM^G$ and $LLM^S$ (Section~\ref{subsection:statistics}). Additionally, we collect the hint candidates $H$, which form part of the input for $LLM^S$, along with the optimal hint $h^*$ in $H$, which serves as the ground truth for $LLM^G$. The index $idx^*$ of $h^*$ is used as the ground truth for $LLM^S$ (Section~\ref{subsection:label_collection}). An example of the input and output for \Model(G) and \Model(S) is shown in Figure ~\ref{fig:input_output}.

\subsection{Construction of Statistics}
\label{subsection:statistics}

\begin{algorithm}[t]
\caption{Obatin\_Statistics}
\label{alg:obtain_stat}
\begin{algorithmic}[1] 
\Require query $q$, database instance $d$
\Ensure related statistics $stats$
\State $Card\_Tb, NDV,Main\_Value, Min\_Max, Hist \gets []*5$
\State $T$ $\gets$ tables related to $q$
\State $p_d$ $\gets$ default plan searched by DBMS in $d$

\For{each table $t$ in $T$}
\State Append $card$ in $p_d$ and rows numbers of $t$ to $Card\_Tb$
    \For{each column $col$ in $t$}
        \State Append ndv of $col$ to $NDV$
        \State Append main value of $col$ to $Main\_Value$
        
        \If{$col$ is numerical}
        \State Append min and max values of $col$ to $Min\_Max$
        \State Append histograms of $col$ to $Hist$
        \EndIf
    \EndFor
\EndFor
\State $stats$ $\gets$ $\{Card\_Tb, NDV, Main\_Value, Min\_Max, Hist\}$
\State \Return $stats$
\end{algorithmic}
\end{algorithm}

        


We define and describe the collection of $stats$ in this part. Since it serves as external knowledge to bridge the gap between LLMs and conventional optimizers, $stats$ is utilized as a fundamental part of the LLM input. In \Model, $stats(d,q) $ consists of \{$ Card\_Tb$, $NDV$, $Main\_Value$, 
$Min\_Mas$, $Hist$ \}, which are essential for traditional optimizers. The procedure for collecting these statistics is outlined in Algorithm ~\ref{alg:obtain_stat}, with detailed definitions of each term provided below. 


\vpara{(1) Card\_Tb} represents the DBMS estimated cardinality and total row count for each base table $t$. Specifically, cardinality represents estimated rows for each node in the query plan. As it highly influences the cost in joining relations, we consider $card$ of base tables a valuable feature for learning. 
Specifically, we extract cardinality from the default execution plan $p_d$ searched by PG. The default plan $p_d$ can be obtained by running the $\mathsf{EXPLAIN}$ command without actual execution. Additionally, we consider the total number of rows $|t|$ of each table $t$ to be a useful feature for the LLM. Therefore, we formally present this term as $\{t:card(|t|)\}$.

\vpara{(2) NDV} refers to the \textbf{N}umber of \textbf{D}istinct \textbf{V}alues. For each column $col$ in each table $t$, there exists a corresponding distinct value count, denoted as $ndv$. Since operations such as $\mathsf{DISTINCT}$ or aggregate functions like $\mathsf{Count(\cdot)}$ effectively disregard duplicate values, the number of distinct values for each column remain a crucial factor in query optimization. Thus, we represent this statistic as $\{t .col:ndv\}$. 

\vpara{(3) Main\_Value} represents the most common values $value$ and their corresponding frequencies $frequency$ for each column $col$. We hypothesize that values frequently appearing in tables are also likely to appear in queries, making this statistic valuable for LLMs in generating or selecting the optimal query plan from plan candidates. We incorporate it into $stats$ using the format $\{t.col:[value, frequency]\}$.

\vpara{(4) Min\_Max} represents the minimum and maximum values for each numerical column $col$. For predicate filters such as $col > constant$, it is clear that the query will yield no results if the constant exceeds the maximum value of $col$. A similar logic applies to minimum values. This statistic is formulated as $\{t.col:[min,max]\}$.

\vpara{(5) Hist} represents the histogram of numerical columns calculated by the DBMS. In PG, an equal-height histogram is computed, which reflects the distribution of a single column within database instance $d$.
By using the $\mathsf{ANALYZE}$ command, we can obtain the histogram in the form $hist = [v_1,...,v_k]$ where $v_i$ represent the boundaries of histogram bins and $k$ denotes the total number of bins. In this paper, we formulate the histogram statistic as $\{t.col:hist\}$.






\subsection{Collecting Labels}
\label{subsection:label_collection}

\begin{figure*}
    \centering
    \includegraphics[width=1\linewidth]{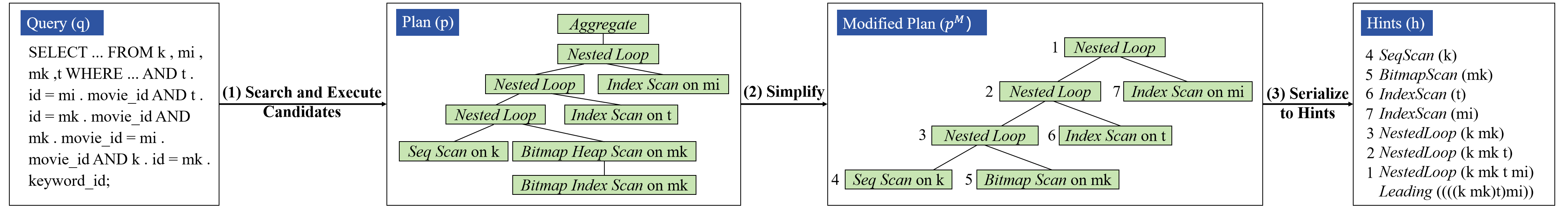}
    \caption{An example of collecting labels for \Model\ by searching for candidates, simplify these plans and serializing plans to sequential hints.}
    \label{fig:PHP}
\end{figure*}

\begin{algorithm}[t]
\caption{Transform\_Plan}
\label{alg:transform_pl}
\begin{algorithmic}[1] 
\Require plan tree $p$
\Ensure sequential hint $h$
\State scan hints $S$, join hints $J$ $\gets []*2$
\State $root \gets$ root of $p$ 
\Function{Traverse}{Node $n$}
\If{$n$ is a leaf node}
\State $ScanType \gets$ scan type of $n$
\State $t \gets$ table of $n$
\State Append ``$ScanType(t)$'' to $S$
\State \Return $t$
\ElsIf{$n$ has two children}
\State $JoinType \gets$ join type of $n$
\State $l, r \gets$ the left child and right child of $n$
\State $LTs \gets$ \Call{Traverse}{$l$}
\State $RTs \gets$ \Call{Traverse}{$r$}
\State Append ``$JoinType(LTs\ RTs)$'' to $J$
\State \Return ($LTs\ RTs$)
\Else
\State $Child \gets$ the child of $n$
\State \Return \Call{Traverse}{$Child$}
\EndIf
\EndFunction
\State $ATs \gets$ \Call{Traverse}{$root$}
\State $L \gets$ ``$Leading(ATs)$''
\State $h \gets \{S,J,L\}$
\State \Return $h$
\end{algorithmic}
\end{algorithm}

Given a query, we collect the hint candidates $H$ as input for $LLM^S$, the optimal hint $h^*$ as the label for $LLM^G$, and the index $idx^*$ of $h^*$ as the label for $LLM^S$. To obtain these, we first search for plan candidates $P$ using multiple \Baoarm s and execute them in a real DBMS to measure their latencies. To enable the sequential input and output required by LLMs, each plan candidate is transformed into a serialized hint sequence, as described in Algorithm ~\ref{alg:transform_pl}, which involves both plan simplification and serialization stages. The hint corresponding to the query plan with the lowest execution latency is identified as the optimal hint $h^*$. Below, we detail the processes of searching and executing candidates, simplifying query plans, and serializing plans into hints.

\vpara{(1) Searching and Executing Candidates.}
We utilize \Baoarm s~\cite{marcus2021bao}, as discussed in Section \ref{subsection:bao}, to search for plan candidates and execute all plans in a real DBMS to evaluate their performance. After executing all plan candidates, the plan with the lowest execution latency is selected as the optimal plan $p^*$. As noted in prior work~\cite{marcus2021bao}, using five \Baoarm s achieves a 93\% improvement in optimization performance. Therefore, in this paper, we use the plans corresponding to these arms as candidate plans $P$ to reduce inference time. 

To handle queries that may take excessive time to execute, we impose a timeout limit of 3 minutes, discarding any queries exceeding this threshold. Furthermore, to minimize the data collection effort for the training set, we terminate plans that take longer to execute than the currently known optimal plan, as they cannot be optimal. An example in Figure~\ref{fig:PHP} illustrates a query with four tables and a searched plan candidate $p$ from $P$.

\vpara{(2) Simplifying Query Plans.}
A query plan is typically represented as a tree, where nodes correspond to physical operations. Generally, the plan tree consists of scan nodes, join nodes, and other nodes such as aggregation or sort operations. Since nodes like aggregation and sort do not influence query optimization, we simplify the query plan into $p^M$, retaining only scan and join nodes, as described in lines 16 to 18 of Algorithm ~\ref{alg:transform_pl}.

For example, in Figure~\ref{fig:PHP}, the query plan $p$ contains a bitmap heap scan and bitmap index scan, which collectively form a bitmap scanning process. These are consolidated into a single bitmap scan node in the simplified plan tree $p^M$. Additionally, we omit the aggregation node, as it pertains only to the \textit{SELECT} statement and remains identical across all plan candidates. After this step, we obtain the simplified plan tree $p^M$.

\vpara{(3) Serializing Plans into Hints.}
After simplifying the query plan to $p^M$, each remaining node in $p^M$ is serialized into either a scan hint in $S$ or a join hint in $J$ using a preorder traversal, as shown in Algorithm ~\ref{alg:transform_pl}. A leading hint $L$ is also generated to define the join order of base relations in $p^M$. By combining these hints, we transform a query plan $p$ into a structured text sequence $h$, which adheres to the input and output requirements of LLMs.

For instance, the simplified plan tree $p^M$ in Figure~\ref{fig:PHP} consists of four scan nodes (leaf nodes) corresponding to four tables and three join nodes (non-leaf nodes) connecting the scan nodes. Each scan node in $p^M$ is converted into a scan hint, and each join node is represented as a join hint in the format specified by \PHP. Hints 4–7 correspond to scan nodes and are generated using a preorder traversal of $p^M$, while hints 1–3 correspond to join nodes and are listed after the scan hints in $h$.

The \textit{Leading} hint specifies the join order of the query plan by listing all tables in the query plan following an in-order traversal of $p^M$. For example, in Figure~\ref{fig:PHP}, node 3 connects nodes 4 and 5, forming \textit{(k mk)}, while node 2 connects node 3 and node 6, yielding \textit{((k mk)t)}. Finally, node 1, as the root, connects node 2 and node 7, resulting in \textit{(((k mk)t)mi)}. Thus, the leading hint $L$ is expressed as \textit{Leading((((k mk)t)mi))}.

\subsection{Fine-Tuning Models}
For our supervised fine-tuning, the dataset for $LLM^G$ is defined as $\mathbb{D^G} = {\{\text{Input: } \langle q, \text{stats}(q, d) \rangle, \text{Output: } \langle h^* \rangle\}}$, and the dataset for $LLM^S$ is defined as $\mathbb{D^S} = {\{\text{Input: } \langle q, \text{stats}(q, d), H \rangle, \text{Output: } \langle \text{idx}^* \rangle\}}$. After collecting the necessary data as described in Section ~\ref{subsection:statistics} and Section ~\ref{subsection:label_collection}, we employ the cross-entropy loss, as outlined in Section ~\ref{subsection:fine-tuning-LLMs}, to fine-tune the LLMs in \Model.
The training objective for $LLM^G$ is to maximize $p_{\theta_G}(h^*|q, \text{stats}(q, d))$, while the training objective for $LLM^S$ is to maximize $p_{\theta_S}(\text{idx}^*|q, \text{stats}(q, d), H)$.




\section{Experiment}
\label{section:experiment}
In this section, we present an experimental study that evaluates \Model from the following perspectives:
\begin{itemize}[leftmargin=1em]
    \item \textbf{Main Evaluation:} We directly evaluate \Model(G) and \Model(S) as standalone approaches in terms of execution latency. (Section ~\ref{subsection:main_res})

    \item \textbf{End-to-End Latency:} We assess the end-to-end latency of \Model(G) and \Model(S), as this metric is critical in real-world scenarios. (Section ~\ref{subsection:end2end})
    
    \item \textbf{Independent Component Performance:} Since $LLM^G$ and $LLM^S$ serve as the candidate generator and candidate selector in \Model(G) and \Model(S), respectively, we evaluate their independent performance without interference from other components. (Sections ~\ref{subsection:generative_model_potential} and ~\ref{subsection:select_model_evaluation})

    \item \textbf{Combined Approach:} We combine \Model(G) and \Model(S) to form a unified approach, \Model(G+S), and evaluate its combined performance. (Section ~\ref{subsection:g+s})
    
    \item \textbf{Ablation Study:} We analyze the impact of various factors on the performance of \Model, including input statistics, inference strategies, backbones, and fine-tuning techniques. (Section ~\ref{subsection:ablation_study})
    
\end{itemize}


\subsection{Experimental Settings}
\vpara{Database engine and Environment.}
We use PostgreSQL 16.1 along with the corresponding version of \PHP\ as the experimental environment for \Model. \PHP\ is modified as described in Section ~\ref{subsection:pg_hint_plan}. The database operates on a server equipped with an Intel Xeon Gold 5218 CPU (40 cores, 80 threads) and 251GB of RAM. Fine-tuning of the LLMs is conducted on a separate machine featuring an Intel Xeon Platinum 8358 CPU, 960GB of RAM, and 4 NVIDIA A800 GPUs.

\vpara{Databases.}
We evaluate \Model\ on three benchmarks.
\begin{itemize}[leftmargin=1em]
    \item\textbf{Join-Order-Benchmark (JOB)} is a benchmark based on a real-world database \textbf{IMDb} related to internet movies. IMDb has a total size of 3.6GB (8.5GB including indexes), containing data about various movies, directors and actors, distributed across 21 tables. JOB consists of 113 queries, derived from 33 templates, with the number of tables per query ranging from 4 to 17.

    \item \textbf{JOB-EXT} is built on JOB and uses the same database, IMDb. It extends JOB with 24 additional queries derived from 14 different templates, covering queries that involve 3 to 11 tables.
    \item \textbf{Stack} is a real-world dataset derived from questions and answers on StackExchange websites~\cite{marcus2021bao}. The dataset contains 10 tables and spans 100GB. Stack consists of 11 templates, covering queries involving 4 to 12 tables. 
\end{itemize}

    
\vpara{Baselines.} We evaluate \Model\ against three state-of-the-art query optimization baselines~\cite{marcus2021bao, yu2022cost} as demonstrated in recent studies~\cite{lehmann2023your}. 
\begin{itemize}[leftmargin=1em]
    \item \textbf{PostgreSQL (PG)} is a well-known DBMS, commonly used in previous query optimization studies. In this paper, we utilize its optimizer, referred to as PG, in the following experiments. Based on recommendations from an expert administrator and our hardware environment, we configure PG 16.1 with the following settings: a cache size of 192GB, maintenance working memory of 2GB, WAL buffers of 16MB, effective I/O concurrency set to 2, work memory of 82MB, a WAL size ranging from 1GB to 4GB, a maximum of 40 worker processes, a maximum of 4 parallel workers per gather, a total of 40 parallel workers and up to 4 parallel maintenance workers. All experiments in the paper are conducted in these settings.
    
    \item \textbf{BAO~\cite{marcus2021bao}} utilizes the optimizer in PG to select the optimal plan by choosing the optimal \Baoarm\ with a Tree-CNN, as detailed in section ~\ref{subsection:bao}. We utilize the open-source code of BAO and train its Tree-CNN with all plan-latency pairs collected in Section ~\ref{subsection:label_collection}. This setup ensures a fair comparison between BAO and \Model.
    
    \item \textbf{HybridQO~\cite{yu2022cost}} follows a similar approach to BAO. It employs MCTS to explore plan candidates and selects the optimal one using a Tree-LSTM. We follow its open-source code and train both MCTS and Tree-LSTM with the plan-latency pairs.
\end{itemize}

\vpara{Evaluation Metrics.} 
We evaluate \Model\ from several aspects using metrics such as execution latency, end-to-end latency, and the statistical properties (minimum, average, and standard deviation) of latencies for multiple samples, as well as selection accuracy.

\textbf{Execution Latency.}
To compare \Model\ with other methods, we use total execution latency ($Exec$), a widely adopted metric in prior works~\cite{chen2023leon,marcus2021bao,marcus2019neo,lehmann2023your,zhu2023lero,yang2022balsa}. Additionally, we report tail latencies to assess \Model's stability. These metrics are primarily used in Section ~\ref{subsection:main_res}, Section ~\ref{subsection:g+s}, and Section ~\ref{subsection:ablation_study} for the main evaluation, combined approach evaluation, and ablation studies.

\textbf{End-to-End Latency.}
End-to-end latency ($E2E$) measures the total time from query reception to execution completion, making it a critical metric for real-world applications. We use this metric to evaluate the end-to-end performance of \Model(G) and \Model(S) in Section ~\ref{subsection:end2end}. Specifically, $E2E$ includes the time for statistic collection, LLM inference, DBMS planning, and execution. For \Model(G), additional latency arises from the Tree-CNN selection process. For \Model(S), $E2E$ also accounts for the time required to collect candidate plans and transform plans into hints.

\textbf{Metrics for Candidate Generation.}
To evaluate candidate generation methods such as $LLM^G$, we introduce several metrics. For each query, a set of plan candidates is generated, and their performance is evaluated using the minimum execution latency, average execution latency, and the standard deviation of execution latencies. These metrics, denoted as $Min$, $Avg$, and $Std$, are summed across all queries and are used in Section ~\ref{subsection:generative_model_potential}.

\textbf{Selection Accuracy.}
To assess the performance of the list-wise $LLM^S$, we use selection accuracy, which measures the percentage of queries for which the model identifies the optimal plan. This metric is employed in Section ~\ref{subsection:select_model_evaluation}. Other metrics, such as Q-error, are not applicable to a list-wise cost model and are therefore excluded from our evaluation.

\vpara{Implementation Settings.} 
We configure several implementation settings related to query construction, query execution, training configuration, and inference refinement:

\textbf{Query Construction.}
We apply different query construction strategies for the JOB, JOB-EXT, and Stack benchmarks:
For \textbf{JOB and JOB-EXT}, these benchmarks provide templates for query generation. We use the provided templates to construct our test set and prepare additional queries using an existing method~\cite{negi2023robust}. After preparation, the queries are randomly split into a training set and a validation set of 100 queries. The training set contains 8,592 queries for JOB and 9,722 queries for JOB-EXT. 
For \textbf{Stack}, this benchmark provides a sufficient query set covering 11 templates. Thus, no additional queries are prepared. For each template, the queries are randomly divided into a training-validation set and a test set. The training-validation set is further split into a training set of 4,850 queries and a validation set of 100 queries. The final test set for Stack contains 76 queries.

\textbf{Query Execution.}
Due to limitations in \PHP, hints for nested queries cannot be correctly applied to the DBMS. Therefore, \Model\ does not process nested queries.
To mitigate cold cache effects, we follow recent recommendations~\cite{lehmann2023your} and measure each query's latency after performing two warm-up executions before the final execution.
To avoid unpredictable latency issues, a timeout of 3 minutes is set for both data collection and evaluation. With this setting, data collection takes approximately 9 hours for JOB, 17 hours for JOB-EXT, and 19 hours for Stack.

\textbf{Training Configuration.}
We use Llama3.1-8B-nstruct~\cite{dubey2024llama}, a powerful open-source PLM, as the backbone for \Model.
For JOB and JOB-EXT, the base model is fine-tuned with a batch size of 64 and a learning rate of 5e-6. For Stack, due to the smaller size of the training data, we use a batch size of 32 while keeping the learning rate at 5e-6.
Both $LLM^G$ and $LLM^S$ are trained for 8 epochs on each benchmark, and their performance on the validation set is used to select the best epoch for reporting. Each training process takes approximately 10 hours.

\textbf{Inference Refinement.}
To handle invalid outputs produced by LLMs, we replace invalid outputs with PG plan representations. Invalid outputs include incorrectly formatted hints from $LLM^G$ or out-of-range indices from $LLM^S$.
In practice, $LLM^S$ produces no invalid outputs, while approximately 5\% of $LLM^G$ outputs are invalid.




\subsection{Main Results}

\label{subsection:main_res}
In this section, we evaluate \Model(G) and \Model(S) as shown in Table ~\ref{tab:main_res}.

Overall, both \Model(G) and \Model(S) outperform PG, BAO, and HybirdQO across all benchmarks. Specifically, \Model(G) achieves an improvement of 67.2\% ($\frac{129.26-42.42}{129.26} =67.2\%$) on JOB, 
29.6\% ($\frac{66.41-46.72}{66.41} =29.6\%$) on JOB-EXT,
and 17.4\% ($\frac{334.66-276.54}{334.66} =17.4\%$) on Stack compared to PG. 
Furthermore, replacing the cost model in BAO with an LLM results in an improvement of 48.4\% ($\frac{70.68-36.45}{70.68} =48.4\%$) on JOB, 
6.1\% ($\frac{61.77-58.02}{61.77} =6.1\%$) on JOB-EXT, 
and 45.8\% ($\frac{405.71-219.86}{405.71} =45.8\%$) on Stack compared to BAO. 

\begin{figure*}
    \centering
    \includegraphics[width=1\linewidth]{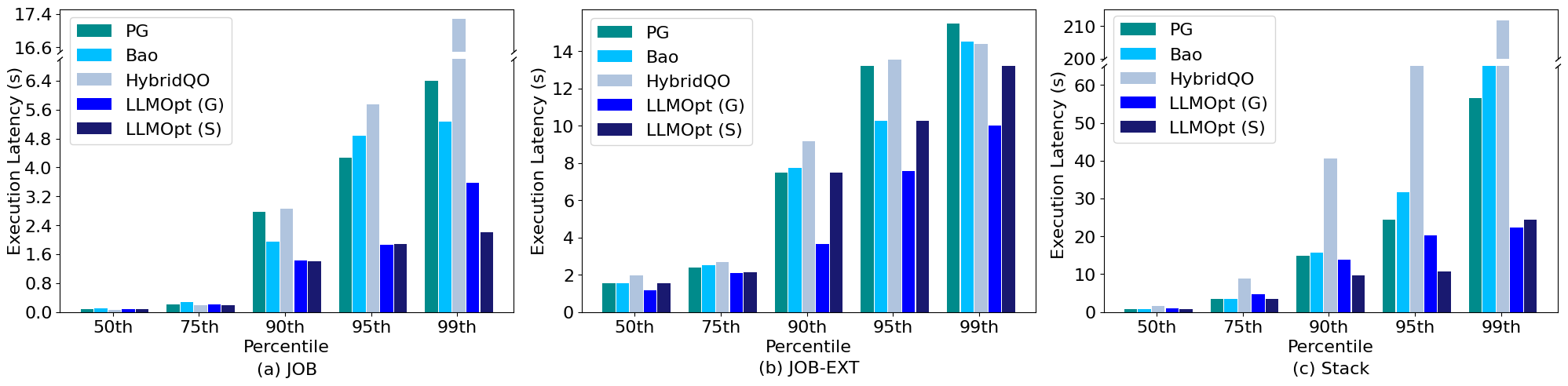}
    \caption{The 50\%, 75\%, 90\%, 95\% and 99\% percentiles of execution latency in PG, BAO, HybridQO, \Model(G) and \Model(S).}
    \label{fig:percentiles}
\end{figure*}

A single query with high latency can significantly affect the overall execution time, even when other queries perform well. Thus, we compare the tail latencies at 50\%, 75\%, 90\%, 95\% and 99\% percentiles for each method. The results, shown in Figure ~\ref{fig:percentiles}, indicate that \Model(G) and \Model(S) outperform all baselines across all workloads. Specifically, \Model(S) achieves the best performance on JOB and most percentiles on Stack while \Model(G) performs better on JOB-EXT. These percentile results explain the ranking pattern observed in Table ~\ref{tab:main_res}.

This finding highlights that LLMs are effective either at generating an optimal plan or selecting an optimal plan. Notably, \Model(S) performs better than \Model(G) on the JOB and Stack benchmarks, while \Model(G) achieves better performance on JOB-EXT. In Section ~\ref{subsection:generative_model_potential} and Section ~\ref{subsection:select_model_evaluation}, we explore the performance of $LLM^G$ in \Model(G) and $LLM^S$ in \Model(S).

\begin{table}[t]
\centering
\small
\caption{Execution latencies $Exec$ (s) $\downarrow$ and end-to-end latencies $E2E$ (s) $\downarrow$.}
\label{tab:main_res}
\begin{tabular}{ccccccc}
\toprule
 & \multicolumn{2}{c}{JOB} & \multicolumn{2}{c}{JOB-EXT} & \multicolumn{2}{c}{Stack} \\
 & $Exec$ & $E2E$ & $Exec$ & $E2E$ & $Exec$ & $E2E$ \\
\midrule
PostgreSQL & 129.26& 129.34& 66.41 & 66.55 & 334.66 & \underline{335.62}\\
BAO\cite{marcus2021bao} & 70.68 & \underline{75.88} & 61.77 & 62.72 &  405.71 &  410.38\\
HybridQO\cite{yu2022cost} &175.56 &183.87 & 83.72 & 84.62 & 1084.91 & 1097.7 \\  
\Model (G) & \underline{42.42}  & 148.78 & \textbf{45.22} & \textbf{55.98} & \underline{300.50} &385.07 \\  
\Model (S) & \textbf{36.45} & \textbf{62.51} & \underline{58.02} & \underline{61.07}  & \textbf{219.86} & \textbf{236.15} \\   
\bottomrule
\end{tabular}
\end{table}

\subsection{End-to-End Latency Evaluation}
\label{subsection:end2end}



We evaluate end-to-end latency in this section as shown in Table ~\ref{tab:main_res}. Overall, \Model(G) performs worse and \Model(S) still outperforms other baselines. 

The main reason for this difference is their inference cost. \Model(G) spends an excessive amount of time generating multiple complete plans consisting of  all $\{S,J,L\}$, whereas \Model(S) only generates a single token representing the optimal index in $H$. Since transformers, the fundamental architecture of LLMs, generate output token-by-token, \Model(G) takes significantly longer to complete inference. As a result, its fails to outperform baselines in terms of end-to-end latency.  

In conclusion, \Model(S) proves to be the most practical choice, as it achieves the lowest end-to-end latency among all baselines across all benchmarks.  

\subsection{Evaluation of $LLM^G$}
\label{subsection:generative_model_potential}
\begin{table}[t]
\centering
\caption{Minimum (s) $\downarrow$, average (s) $\downarrow$, and standard deviation (s) of execution latencies for multiple generated plans.}
\label{tab:mean_std}

\resizebox{0.5\textwidth}{!}{
\begin{tabular}{ccccccc}
\toprule
 & \multicolumn{1}{c}{JOB} &\multicolumn{1}{c}{JOB-EXT}& \multicolumn{1}{c}{Stack} \\
 &$Min$ /$Avg$ / $Std$ &$Min$ / $Avg$ / $Std$ &$Min$ /$Avg$ /$Std$ \\
\midrule
\Baoarm s\cite{marcus2021bao} & {35.62} / {118.88} / {86.56} & {49.46} / 107.42 / {59.50} &{210.92} / 516.34 / {328.01}  \\
MCTS\cite{yu2022cost}&96.46 /1,287.97/ 832.47& 55.53 / 259.32/ 199.80 & 332.58 / 1,141.06 / 376.12\\
$LLM^G$ & \textbf{33.74} / \textbf{53.68} / \textbf{16.05} & \textbf{45.09} / \textbf{84.46} / \textbf{30.07}  & \textbf{202.88} / \textbf{422.63} / \textbf{295.26}
 \\
\bottomrule
\end{tabular}}
\end{table}

Following the scenario described in Section ~\ref{subsection:main_res}, where $LLM^G$ generates hint candidates corresponding to the plan candidates, we now focus on evaluating the performance of $LLM^G$ itself without the influence of the cost model. Other heuristic methods, such as \Baoarm s in BAO~\cite{marcus2021bao} and MCTS~\cite{yu2022cost} in HybridQO, also search for a candidate plan set. In this section, we assess $LLM^G$'s ability from two perspectives: stability and its maximal potential.

\vpara{Stability.} In this aspect, we evaluate the stability of $LLM^G$ by analyzing the average and the standard deviation of execution latencies.
The results presented in Table ~\ref{tab:mean_std} show that \Model(G) achieves a lower mean value and standard deviation. This demonstrates that it is more stable than \Baoarm s and MCTS. Even if a less effective selection method is used to choose query plans, \Model(G) maintains good performance in terms of execution latency. 

\vpara{Maximal Potential.}
In this regard, we evaluate the maximal potential of $LLM^G$. Without using a cost model, the optimal execution latency in the plan candidates serves as another metric for evaluating the set's quality.
Assuming the cost model is 100\% accurate (i.e. always able to select the optimal plan from the given candidate set), we assess the performance of the generative model as shown in Table ~\ref{tab:mean_std}. 
The results indicate that $LLM^G$ achieves better performance than all other methods across all workloads, highlighting the strong generation capability of LLMs. 

The evaluations above demonstrate the comprehensive capabilities of LLM as a generating model: ensuring high stability and excellent potential in generating optimal plans.

\begin{table*}[h]

    \centering
    \caption{Ablation Study of \Model\ on JOB in terms of Execution Latency (s) $\downarrow$ and tail latency (s) $\downarrow$. }
    \label{tab:ablation_study}
    \begin{tabular}{@{}l|cccccc|cccccc}
        \hline 
          & \multicolumn{6}{c|}{\Model(G)} & \multicolumn{6}{c}{\Model(S)}\\
        Variants & $Exec$ & 50th & 75th & 90th & 95th & 99th &$Exec$ & 50th & 75th & 90th & 95th & 99th\\
        \cmidrule(r){1-1}\cmidrule(r){2-7}\cmidrule(r){8-13}
        \Model& 42.42 &0.078&0.218&1.433&1.860&3.567 & 36.45 & 0.072& 0.185&1.407 & 1.883& 2.207\\  
        \cmidrule(r){1-1}\cmidrule(r){2-7}\cmidrule(r){8-13}
        Input Statistics& & & & & & & & & & & &\\  
        \quad- w/o ALL Statistics  & 56.66 &0.075&0.223&1.975&3.227&5.613&46.86&0.072&0.185&1.604&2.207&3.445\\
        \quad- w/o $Card\_Tb$& 51.00 & 0.094&0.204&1.234&2.145&4.492  & 41.95& 0.072& 0.185& 1.422&2.019 & 3.320\\
        \quad- w/o $NDV$&42.41 &0.085&0.182&1.683&2.176&2.683  & 36.84&0.072&0.215&1.314&1.728&2.132\\
        \quad- w/o $Main\_Value$ & 43.63 & 0.084&0.192&1.642&2.062&3.471 & 35.29 & 0.072&0.185&1.314&1.728&2.132\\
        \quad- w/o $Min\_Max$&  47.75 &0.086&0.183&1.540&2.344&4.320  & 41.45 & 0.072&0.191&1.407&1.883&2.984 \\
        \quad- w/o $Hist$& 47.24 & 0.092&0.224&1.520&2.036&4.842  & 45.63 & 0.072&0.191&1.475&2.064&3.445\\
       
        \cmidrule(r){1-1}\cmidrule(r){2-7}\cmidrule(r){8-13}
        Inference Strategies& & & & & & & & & & & &\\ 
        \quad- Major Voting& 38.30&0.084&0.239&1.289&1.872&2.450 & 36.43 & 0.072&0.191&1.407&1.883&2.207\\
        \quad- Greedy Decoding & 39.25 & 0.088&0.183&1.427&1.874&2.170 &- &- &- &- &- &- \\
        \cmidrule(r){1-1}\cmidrule(r){2-7}\cmidrule(r){8-13}
        Backbones& & & & & & & & & & & &\\ 
        \quad- LlaMA3-8B\cite{dubey2024llama}& 58.45& 0.090&0.228&1.720&3.975&5.398  & 40.11& 0.072&0.185&1.407&1.883&2.207 \\
        \quad- Deepseek-Coder-6.7B\cite{guo2024deepseek} & 63.44 &0.084&0.189&1.563&2.028&3.473 &40.20& 0.072&0.191&1.407&1.883&2.207\\
        \cmidrule(r){1-1}\cmidrule(r){2-7}\cmidrule(r){8-13}
        Prompting PLMs& & & & & & & & & & & &\\ 
        \quad- GPT-4o\cite{hurst2024gpt} & 69.30 & 0.107&0.248&1.816&4.529&5.873& 75.15& 0.134&0.433&2.047&4.002&4.563\\
        \hline
    \end{tabular}
    
\end{table*}

\begin{table}[htbp]
\centering
\caption{Execution latencies (s)  $\downarrow$ of standalone and combination approaches.}
\label{tab:G+S}
\begin{tabular}{cccc}
\toprule
 & JOB & JOB-EXT & Stack \\
\midrule
\Model (G) & 42.42 & \underline{45.22} &300.50 \\
\Model (S) & \textbf{36.45} & 58.02  & \textbf{219.86} \\
\Model(G+S) & \underline{39.52} & \textbf{42.12}  & \underline{263.05} \\
\bottomrule
\end{tabular}
\end{table}
\subsection{Evaluation of $LLM^S$}
\label{subsection:select_model_evaluation}
\begin{table}[t]
\centering
\caption{Selection accuracy (\%) $\uparrow$ of $LLM^S$.}
\label{tab:selection}
\begin{tabular}{cccc}
\toprule
 & JOB & JOB-EXT & Stack \\
\midrule
Tree-CNN\cite{marcus2021bao} & 31.86& 50.00&  46.05\\
Tree-LSTM\cite{yu2022cost} &43.36& 25.00  & 26.32\\
$LLM^S$ & \textbf{76.10} & \textbf{66.66}  & \textbf{63.16}
 \\
\bottomrule
\end{tabular}
\end{table}

In this section, we independently evaluate $LLM^S$ in comparison with other cost models such as Tree-CNN and Tree-LSTM. Since $LLM^S$ functions as a list-wise cost model, we assess its performance based on selection accuracy. The results shown in Table ~\ref{tab:selection} indicate that $LLM^S$ is the most powerful selection method, achieving the highest accuracy among all baselines and workloads. 
This superior performance is due to the use of full collected statistics as part of the input and the excellent understanding ability of LLMs.

\subsection{Evaluation of Combined Approach} 
\label{subsection:g+s}
In \Model(G), LLM is utilized as a candidate generator, and in \Model(S), LLM is utilized as a candidate selector. It is straightforward to combine these approaches into \Model(G+S), a potential solution for query optimization. In this combined approach, $LLM^G$ generates the plan candidates, and $LLM^S$ is used to select the optimal plan from the candidate set. The results of this combined approach are presented in Table~\ref{tab:G+S}.

Ideally, the combination of $LLM^G$ and $LLM^S$ is expected to outperform either individual model, due to the enhanced generation ability and more accurate selection. However, except for the JOB-EXT benchmark, \Model(G+S) performs worse than \Model(S). We believe this performance reduction mainly arises from a distribution shift between the training and inference phases.
Specifically, candidates in the training data for $LLM^S$ are generated by \Baoarm s, while it takes candidates from $LLM^G$ as input in the usage in \Model(G+S). The distribution shift causes the inaccuracy of $LLM^S$ in \Model(G+S), leading to a worse performance. This leave room for future works to refine.


\subsection{Ablation study}
\label{subsection:ablation_study}
We conduct the ablation studies for \Model(G) and \Model(S) and show the results in Table ~\ref{tab:ablation_study}.

\subsubsection{Ablation study on Input Statistics.}
In addition to the origin query $q$, statistics $stats$ are a main component of the input sequence for LLMs. In \Model, $stats$ consists of five key statistics. Our ablation study shows that all statistics generally contribute to improving performance. Specifically, $NDV$ and $Main\_Value$ are less critical, as omitting these statistics results in the smallest performance reduction in the generative approach and the selective approach. The ablation study excluding all statistics further confirms the necessity of relevant statistics in enhancing model performance.

\subsubsection{Ablation Study on Inference Strategies.}
In \Model(G), sampling is used as the inference strategy to generate diverse plan candidates, while in \Model(S), a greedy decoding strategy is employed for selecting the most promising plan. Greedy decoding introduces no randomness in the output, and majority voting involves selecting the most frequent output from the generated plan candidates. We investigate the impact of different inference strategies on performance.

Our experimental results reveal that when Tree-CNN is not used in \Model(G), majority voting and greedy decoding achieve better performances. This aligns with the observation mentioned in Section ~\ref{subsection:main_res}, where Tree-CNN is found to be an ineffective selection method. Despite this, we choose to sample multiple outputs from \Model(G) to explore its potential as described in Section ~\ref{subsection:generative_model_potential}. For \Model(S), greedy decoding outperforms the majority voting strategy, which is choosen as the inference strategy in \Model(S).

\subsubsection{Ablation Study on Model Backbones.}
The base model used in \Model\ is LLaMA3.1-8B. To evaluate the impact of different base models on query optimization performance, we replace the base model with other PLMs, such as LLaMA3-8B and DeepSeek-Coder-6.7B. Replacing the backbone model results in a noticeable performance decline, particularly in $LLM^G$, which relies heavily on strong NLU and NLG capabilities. This highlights the necessity of using a powerful base model.

\subsubsection{Ablation Study on LLM Learning Strategy.}
To highlight the importance of fine-tuning in \Model, we test a tuning-free approach using GPT-4o, a powerful closed-source PLM, with few-shot prompting. After introducing the query optimization task and providing a few examples, GPT-4o shows some ability to generate and select optimal query plans. However, it still underperforms compared to our method. This underscores the critical role of supervised fine-tuning, where domain-specific knowledge and task-specific training are essential for optimal performance.

\section{Conclusion}

This paper investigates two distinct approaches that utilize LLMs for the query optimization task. While existing methods based on a search-select pipeline struggle with challenges like a large search space and low selection accuracy, we introduce \Model, which replaces each component with a supervised fine-tuned LLM. Our experiments demonstrate that both approaches, as well as their combination, deliver excellent performance across various workloads. Notably, \Model(S) achieves superior end-to-end latency, making it well-suited for real-world deployment. Future work includes refining the $LLM^S$ component in \Model(G+S) and exploring inference acceleration techniques to further reduce end-to-end latency.



\bibliographystyle{ACM-Reference-Format}
\bibliography{sample}


\begin{thebibliography}{42}


\ifx \showCODEN    \undefined \def \showCODEN     #1{\unskip}     \fi
\ifx \showDOI      \undefined \def \showDOI       #1{#1}\fi
\ifx \showISBNx    \undefined \def \showISBNx     #1{\unskip}     \fi
\ifx \showISBNxiii \undefined \def \showISBNxiii  #1{\unskip}     \fi
\ifx \showISSN     \undefined \def \showISSN      #1{\unskip}     \fi
\ifx \showLCCN     \undefined \def \showLCCN      #1{\unskip}     \fi
\ifx \shownote     \undefined \def \shownote      #1{#1}          \fi
\ifx \showarticletitle \undefined \def \showarticletitle #1{#1}   \fi
\ifx \showURL      \undefined \def \showURL       {\relax}        \fi
\providecommand\bibfield[2]{#2}
\providecommand\bibinfo[2]{#2}
\providecommand\natexlab[1]{#1}
\providecommand\showeprint[2][]{arXiv:#2}

\bibitem[\protect\citeauthoryear{Chen, Chen, Liang, Liu, Wang, Zeng, Su, and Zheng}{Chen et~al\mbox{.}}{2023}]%
        {chen2023leon}
\bibfield{author}{\bibinfo{person}{Xu Chen}, \bibinfo{person}{Haitian Chen}, \bibinfo{person}{Zibo Liang}, \bibinfo{person}{Shuncheng Liu}, \bibinfo{person}{Jinghong Wang}, \bibinfo{person}{Kai Zeng}, \bibinfo{person}{Han Su}, {and} \bibinfo{person}{Kai Zheng}.} \bibinfo{year}{2023}\natexlab{}.
\newblock \showarticletitle{Leon: A new framework for ml-aided query optimization}.
\newblock \bibinfo{journal}{\emph{Proceedings of the VLDB Endowment}} \bibinfo{volume}{16}, \bibinfo{number}{9} (\bibinfo{year}{2023}), \bibinfo{pages}{2261--2273}.
\newblock


\bibitem[\protect\citeauthoryear{Chew, Bollenbacher, Wenger, Speer, and Kim}{Chew et~al\mbox{.}}{2023}]%
        {chew2023llm}
\bibfield{author}{\bibinfo{person}{Robert Chew}, \bibinfo{person}{John Bollenbacher}, \bibinfo{person}{Michael Wenger}, \bibinfo{person}{Jessica Speer}, {and} \bibinfo{person}{Annice Kim}.} \bibinfo{year}{2023}\natexlab{}.
\newblock \showarticletitle{LLM-assisted content analysis: Using large language models to support deductive coding}.
\newblock \bibinfo{journal}{\emph{arXiv preprint arXiv:2306.14924}} (\bibinfo{year}{2023}).
\newblock


\bibitem[\protect\citeauthoryear{Daull, Bellot, Bruno, Martin, and Murisasco}{Daull et~al\mbox{.}}{2023}]%
        {daull2023complex}
\bibfield{author}{\bibinfo{person}{Xavier Daull}, \bibinfo{person}{Patrice Bellot}, \bibinfo{person}{Emmanuel Bruno}, \bibinfo{person}{Vincent Martin}, {and} \bibinfo{person}{Elisabeth Murisasco}.} \bibinfo{year}{2023}\natexlab{}.
\newblock \showarticletitle{Complex QA and language models hybrid architectures, Survey}.
\newblock \bibinfo{journal}{\emph{arXiv preprint arXiv:2302.09051}} (\bibinfo{year}{2023}).
\newblock


\bibitem[\protect\citeauthoryear{Development and Team}{Development and Team}{2012}]%
        {pghintplan}
\bibfield{author}{\bibinfo{person}{NTT OSS Center~DBMS Development} {and} \bibinfo{person}{Support Team}.} \bibinfo{year}{2012}\natexlab{}.
\newblock \showarticletitle{pg\_hint\_plan Github}.
\newblock  (\bibinfo{year}{2012}).
\newblock
\newblock
\shownote{https://github.com/ossc-db/pg\_hint\_plan.}


\bibitem[\protect\citeauthoryear{Dong, Zhang, Ge, Mao, Gao, Lin, Lou, et~al\mbox{.}}{Dong et~al\mbox{.}}{2023}]%
        {dong2023c3}
\bibfield{author}{\bibinfo{person}{Xuemei Dong}, \bibinfo{person}{Chao Zhang}, \bibinfo{person}{Yuhang Ge}, \bibinfo{person}{Yuren Mao}, \bibinfo{person}{Yunjun Gao}, \bibinfo{person}{Jinshu Lin}, \bibinfo{person}{Dongfang Lou}, {et~al\mbox{.}}} \bibinfo{year}{2023}\natexlab{}.
\newblock \showarticletitle{C3: Zero-shot text-to-sql with chatgpt}.
\newblock \bibinfo{journal}{\emph{arXiv preprint arXiv:2307.07306}} (\bibinfo{year}{2023}).
\newblock


\bibitem[\protect\citeauthoryear{Dubey, Jauhri, Pandey, Kadian, Al-Dahle, Letman, Mathur, Schelten, Yang, Fan, et~al\mbox{.}}{Dubey et~al\mbox{.}}{2024}]%
        {dubey2024llama}
\bibfield{author}{\bibinfo{person}{Abhimanyu Dubey}, \bibinfo{person}{Abhinav Jauhri}, \bibinfo{person}{Abhinav Pandey}, \bibinfo{person}{Abhishek Kadian}, \bibinfo{person}{Ahmad Al-Dahle}, \bibinfo{person}{Aiesha Letman}, \bibinfo{person}{Akhil Mathur}, \bibinfo{person}{Alan Schelten}, \bibinfo{person}{Amy Yang}, \bibinfo{person}{Angela Fan}, {et~al\mbox{.}}} \bibinfo{year}{2024}\natexlab{}.
\newblock \showarticletitle{The llama 3 herd of models}.
\newblock \bibinfo{journal}{\emph{arXiv preprint arXiv:2407.21783}} (\bibinfo{year}{2024}).
\newblock


\bibitem[\protect\citeauthoryear{Floridi and Chiriatti}{Floridi and Chiriatti}{2020}]%
        {floridi2020gpt}
\bibfield{author}{\bibinfo{person}{Luciano Floridi} {and} \bibinfo{person}{Massimo Chiriatti}.} \bibinfo{year}{2020}\natexlab{}.
\newblock \showarticletitle{GPT-3: Its nature, scope, limits, and consequences}.
\newblock \bibinfo{journal}{\emph{Minds and Machines}}  \bibinfo{volume}{30} (\bibinfo{year}{2020}), \bibinfo{pages}{681--694}.
\newblock


\bibitem[\protect\citeauthoryear{Gao, Wang, Li, Sun, Qian, Ding, and Zhou}{Gao et~al\mbox{.}}{2023}]%
        {gao2023text}
\bibfield{author}{\bibinfo{person}{Dawei Gao}, \bibinfo{person}{Haibin Wang}, \bibinfo{person}{Yaliang Li}, \bibinfo{person}{Xiuyu Sun}, \bibinfo{person}{Yichen Qian}, \bibinfo{person}{Bolin Ding}, {and} \bibinfo{person}{Jingren Zhou}.} \bibinfo{year}{2023}\natexlab{}.
\newblock \showarticletitle{Text-to-sql empowered by large language models: A benchmark evaluation}.
\newblock \bibinfo{journal}{\emph{arXiv preprint arXiv:2308.15363}} (\bibinfo{year}{2023}).
\newblock


\bibitem[\protect\citeauthoryear{Giannakouris and Trummer}{Giannakouris and Trummer}{2024}]%
        {giannakouris2024demonstrating}
\bibfield{author}{\bibinfo{person}{Victor Giannakouris} {and} \bibinfo{person}{Immanuel Trummer}.} \bibinfo{year}{2024}\natexlab{}.
\newblock \showarticletitle{Demonstrating $\lambda$-tune: Exploiting large language models for workload-adaptive database system tuning}. In \bibinfo{booktitle}{\emph{Companion of the 2024 International Conference on Management of Data}}. \bibinfo{pages}{508--511}.
\newblock


\bibitem[\protect\citeauthoryear{Guo, Yang, Zhang, Song, Zhang, Xu, Zhu, Ma, Wang, Bi, et~al\mbox{.}}{Guo et~al\mbox{.}}{2025}]%
        {guo2025deepseek}
\bibfield{author}{\bibinfo{person}{Daya Guo}, \bibinfo{person}{Dejian Yang}, \bibinfo{person}{Haowei Zhang}, \bibinfo{person}{Junxiao Song}, \bibinfo{person}{Ruoyu Zhang}, \bibinfo{person}{Runxin Xu}, \bibinfo{person}{Qihao Zhu}, \bibinfo{person}{Shirong Ma}, \bibinfo{person}{Peiyi Wang}, \bibinfo{person}{Xiao Bi}, {et~al\mbox{.}}} \bibinfo{year}{2025}\natexlab{}.
\newblock \showarticletitle{Deepseek-r1: Incentivizing reasoning capability in llms via reinforcement learning}.
\newblock \bibinfo{journal}{\emph{arXiv preprint arXiv:2501.12948}} (\bibinfo{year}{2025}).
\newblock


\bibitem[\protect\citeauthoryear{Guo, Zhu, Yang, Xie, Dong, Zhang, Chen, Bi, Wu, Li, et~al\mbox{.}}{Guo et~al\mbox{.}}{2024}]%
        {guo2024deepseek}
\bibfield{author}{\bibinfo{person}{Daya Guo}, \bibinfo{person}{Qihao Zhu}, \bibinfo{person}{Dejian Yang}, \bibinfo{person}{Zhenda Xie}, \bibinfo{person}{Kai Dong}, \bibinfo{person}{Wentao Zhang}, \bibinfo{person}{Guanting Chen}, \bibinfo{person}{Xiao Bi}, \bibinfo{person}{Yu Wu}, \bibinfo{person}{YK Li}, {et~al\mbox{.}}} \bibinfo{year}{2024}\natexlab{}.
\newblock \showarticletitle{DeepSeek-Coder: When the Large Language Model Meets Programming--The Rise of Code Intelligence}.
\newblock \bibinfo{journal}{\emph{arXiv preprint arXiv:2401.14196}} (\bibinfo{year}{2024}).
\newblock


\bibitem[\protect\citeauthoryear{Huang, Li, Zhang, Zhao, Yao, Li, Yu, Zhang, Chen, and Li}{Huang et~al\mbox{.}}{2024}]%
        {huang2024llmtune}
\bibfield{author}{\bibinfo{person}{Xinmei Huang}, \bibinfo{person}{Haoyang Li}, \bibinfo{person}{Jing Zhang}, \bibinfo{person}{Xinxin Zhao}, \bibinfo{person}{Zhiming Yao}, \bibinfo{person}{Yiyan Li}, \bibinfo{person}{Zhuohao Yu}, \bibinfo{person}{Tieying Zhang}, \bibinfo{person}{Hong Chen}, {and} \bibinfo{person}{Cuiping Li}.} \bibinfo{year}{2024}\natexlab{}.
\newblock \showarticletitle{LLMTune: Accelerate Database Knob Tuning with Large Language Models}.
\newblock \bibinfo{journal}{\emph{arXiv preprint arXiv:2404.11581}} (\bibinfo{year}{2024}).
\newblock


\bibitem[\protect\citeauthoryear{Hurst, Lerer, Goucher, Perelman, Ramesh, Clark, Ostrow, Welihinda, Hayes, Radford, et~al\mbox{.}}{Hurst et~al\mbox{.}}{2024}]%
        {hurst2024gpt}
\bibfield{author}{\bibinfo{person}{Aaron Hurst}, \bibinfo{person}{Adam Lerer}, \bibinfo{person}{Adam~P Goucher}, \bibinfo{person}{Adam Perelman}, \bibinfo{person}{Aditya Ramesh}, \bibinfo{person}{Aidan Clark}, \bibinfo{person}{AJ Ostrow}, \bibinfo{person}{Akila Welihinda}, \bibinfo{person}{Alan Hayes}, \bibinfo{person}{Alec Radford}, {et~al\mbox{.}}} \bibinfo{year}{2024}\natexlab{}.
\newblock \showarticletitle{Gpt-4o system card}.
\newblock \bibinfo{journal}{\emph{arXiv preprint arXiv:2410.21276}} (\bibinfo{year}{2024}).
\newblock


\bibitem[\protect\citeauthoryear{Imani, Du, and Shrivastava}{Imani et~al\mbox{.}}{2023}]%
        {imani2023mathprompter}
\bibfield{author}{\bibinfo{person}{Shima Imani}, \bibinfo{person}{Liang Du}, {and} \bibinfo{person}{Harsh Shrivastava}.} \bibinfo{year}{2023}\natexlab{}.
\newblock \showarticletitle{Mathprompter: Mathematical reasoning using large language models}.
\newblock \bibinfo{journal}{\emph{arXiv preprint arXiv:2303.05398}} (\bibinfo{year}{2023}).
\newblock


\bibitem[\protect\citeauthoryear{Jiang, Sablayrolles, Mensch, Bamford, Chaplot, Casas, Bressand, Lengyel, Lample, Saulnier, et~al\mbox{.}}{Jiang et~al\mbox{.}}{2023}]%
        {jiang2023mistral}
\bibfield{author}{\bibinfo{person}{Albert~Q Jiang}, \bibinfo{person}{Alexandre Sablayrolles}, \bibinfo{person}{Arthur Mensch}, \bibinfo{person}{Chris Bamford}, \bibinfo{person}{Devendra~Singh Chaplot}, \bibinfo{person}{Diego de~las Casas}, \bibinfo{person}{Florian Bressand}, \bibinfo{person}{Gianna Lengyel}, \bibinfo{person}{Guillaume Lample}, \bibinfo{person}{Lucile Saulnier}, {et~al\mbox{.}}} \bibinfo{year}{2023}\natexlab{}.
\newblock \showarticletitle{Mistral 7B}.
\newblock \bibinfo{journal}{\emph{arXiv preprint arXiv:2310.06825}} (\bibinfo{year}{2023}).
\newblock


\bibitem[\protect\citeauthoryear{Jiang, Wang, Shen, Kim, and Kim}{Jiang et~al\mbox{.}}{2024}]%
        {jiang2024survey}
\bibfield{author}{\bibinfo{person}{Juyong Jiang}, \bibinfo{person}{Fan Wang}, \bibinfo{person}{Jiasi Shen}, \bibinfo{person}{Sungju Kim}, {and} \bibinfo{person}{Sunghun Kim}.} \bibinfo{year}{2024}\natexlab{}.
\newblock \showarticletitle{A Survey on Large Language Models for Code Generation}.
\newblock \bibinfo{journal}{\emph{arXiv preprint arXiv:2406.00515}} (\bibinfo{year}{2024}).
\newblock


\bibitem[\protect\citeauthoryear{Lagler, Schindelegger, B{\"o}hm, Kr{\'a}sn{\'a}, and Nilsson}{Lagler et~al\mbox{.}}{2013}]%
        {lagler2013gpt2}
\bibfield{author}{\bibinfo{person}{Klemens Lagler}, \bibinfo{person}{Michael Schindelegger}, \bibinfo{person}{Johannes B{\"o}hm}, \bibinfo{person}{Hana Kr{\'a}sn{\'a}}, {and} \bibinfo{person}{Tobias Nilsson}.} \bibinfo{year}{2013}\natexlab{}.
\newblock \showarticletitle{GPT2: Empirical slant delay model for radio space geodetic techniques}.
\newblock \bibinfo{journal}{\emph{Geophysical research letters}} \bibinfo{volume}{40}, \bibinfo{number}{6} (\bibinfo{year}{2013}), \bibinfo{pages}{1069--1073}.
\newblock


\bibitem[\protect\citeauthoryear{Lao, Wang, Li, Wang, Zhang, Cheng, Chen, Tang, and Wang}{Lao et~al\mbox{.}}{2023}]%
        {lao2023gptuner}
\bibfield{author}{\bibinfo{person}{Jiale Lao}, \bibinfo{person}{Yibo Wang}, \bibinfo{person}{Yufei Li}, \bibinfo{person}{Jianping Wang}, \bibinfo{person}{Yunjia Zhang}, \bibinfo{person}{Zhiyuan Cheng}, \bibinfo{person}{Wanghu Chen}, \bibinfo{person}{Mingjie Tang}, {and} \bibinfo{person}{Jianguo Wang}.} \bibinfo{year}{2023}\natexlab{}.
\newblock \showarticletitle{Gptuner: A manual-reading database tuning system via gpt-guided bayesian optimization}.
\newblock \bibinfo{journal}{\emph{arXiv preprint arXiv:2311.03157}} (\bibinfo{year}{2023}).
\newblock


\bibitem[\protect\citeauthoryear{Lehmann, Sulimov, and Stockinger}{Lehmann et~al\mbox{.}}{2023}]%
        {lehmann2023your}
\bibfield{author}{\bibinfo{person}{Claude Lehmann}, \bibinfo{person}{Pavel Sulimov}, {and} \bibinfo{person}{Kurt Stockinger}.} \bibinfo{year}{2023}\natexlab{}.
\newblock \showarticletitle{Is Your Learned Query Optimizer Behaving As You Expect? A Machine Learning Perspective}.
\newblock \bibinfo{journal}{\emph{arXiv preprint arXiv:2309.01551}} (\bibinfo{year}{2023}).
\newblock


\bibitem[\protect\citeauthoryear{Li, Zhang, Liu, Fan, Zhang, Zhu, Wei, Pan, Li, and Chen}{Li et~al\mbox{.}}{2024}]%
        {li2024codes}
\bibfield{author}{\bibinfo{person}{Haoyang Li}, \bibinfo{person}{Jing Zhang}, \bibinfo{person}{Hanbing Liu}, \bibinfo{person}{Ju Fan}, \bibinfo{person}{Xiaokang Zhang}, \bibinfo{person}{Jun Zhu}, \bibinfo{person}{Renjie Wei}, \bibinfo{person}{Hongyan Pan}, \bibinfo{person}{Cuiping Li}, {and} \bibinfo{person}{Hong Chen}.} \bibinfo{year}{2024}\natexlab{}.
\newblock \showarticletitle{Codes: Towards building open-source language models for text-to-sql}.
\newblock \bibinfo{journal}{\emph{Proceedings of the ACM on Management of Data}} \bibinfo{volume}{2}, \bibinfo{number}{3} (\bibinfo{year}{2024}), \bibinfo{pages}{1--28}.
\newblock


\bibitem[\protect\citeauthoryear{Li, Allal, Zi, Muennighoff, Kocetkov, Mou, Marone, Akiki, Li, Chim, et~al\mbox{.}}{Li et~al\mbox{.}}{2023}]%
        {li2023starcoder}
\bibfield{author}{\bibinfo{person}{Raymond Li}, \bibinfo{person}{Loubna~Ben Allal}, \bibinfo{person}{Yangtian Zi}, \bibinfo{person}{Niklas Muennighoff}, \bibinfo{person}{Denis Kocetkov}, \bibinfo{person}{Chenghao Mou}, \bibinfo{person}{Marc Marone}, \bibinfo{person}{Christopher Akiki}, \bibinfo{person}{Jia Li}, \bibinfo{person}{Jenny Chim}, {et~al\mbox{.}}} \bibinfo{year}{2023}\natexlab{}.
\newblock \showarticletitle{Starcoder: may the source be with you!}
\newblock \bibinfo{journal}{\emph{arXiv preprint arXiv:2305.06161}} (\bibinfo{year}{2023}).
\newblock


\bibitem[\protect\citeauthoryear{Liu, Feng, Xue, Wang, Wu, Lu, Zhao, Deng, Zhang, Ruan, et~al\mbox{.}}{Liu et~al\mbox{.}}{2024}]%
        {liu2024deepseek}
\bibfield{author}{\bibinfo{person}{Aixin Liu}, \bibinfo{person}{Bei Feng}, \bibinfo{person}{Bing Xue}, \bibinfo{person}{Bingxuan Wang}, \bibinfo{person}{Bochao Wu}, \bibinfo{person}{Chengda Lu}, \bibinfo{person}{Chenggang Zhao}, \bibinfo{person}{Chengqi Deng}, \bibinfo{person}{Chenyu Zhang}, \bibinfo{person}{Chong Ruan}, {et~al\mbox{.}}} \bibinfo{year}{2024}\natexlab{}.
\newblock \showarticletitle{Deepseek-v3 technical report}.
\newblock \bibinfo{journal}{\emph{arXiv preprint arXiv:2412.19437}} (\bibinfo{year}{2024}).
\newblock


\bibitem[\protect\citeauthoryear{Liu and Mozafari}{Liu and Mozafari}{2024}]%
        {liu2024query}
\bibfield{author}{\bibinfo{person}{Jie Liu} {and} \bibinfo{person}{Barzan Mozafari}.} \bibinfo{year}{2024}\natexlab{}.
\newblock \showarticletitle{Query Rewriting via Large Language Models}.
\newblock \bibinfo{journal}{\emph{arXiv preprint arXiv:2403.09060}} (\bibinfo{year}{2024}).
\newblock


\bibitem[\protect\citeauthoryear{Ma, Gong, He, Zhao, and Duan}{Ma et~al\mbox{.}}{2023}]%
        {ma2023query}
\bibfield{author}{\bibinfo{person}{Xinbei Ma}, \bibinfo{person}{Yeyun Gong}, \bibinfo{person}{Pengcheng He}, \bibinfo{person}{Hai Zhao}, {and} \bibinfo{person}{Nan Duan}.} \bibinfo{year}{2023}\natexlab{}.
\newblock \showarticletitle{Query rewriting for retrieval-augmented large language models}.
\newblock \bibinfo{journal}{\emph{arXiv preprint arXiv:2305.14283}} (\bibinfo{year}{2023}).
\newblock


\bibitem[\protect\citeauthoryear{Marcus, Negi, Mao, Tatbul, Alizadeh, and Kraska}{Marcus et~al\mbox{.}}{2021}]%
        {marcus2021bao}
\bibfield{author}{\bibinfo{person}{Ryan Marcus}, \bibinfo{person}{Parimarjan Negi}, \bibinfo{person}{Hongzi Mao}, \bibinfo{person}{Nesime Tatbul}, \bibinfo{person}{Mohammad Alizadeh}, {and} \bibinfo{person}{Tim Kraska}.} \bibinfo{year}{2021}\natexlab{}.
\newblock \showarticletitle{Bao: Making learned query optimization practical}. In \bibinfo{booktitle}{\emph{Proceedings of the 2021 International Conference on Management of Data}}. \bibinfo{pages}{1275--1288}.
\newblock


\bibitem[\protect\citeauthoryear{Marcus, Negi, Mao, Zhang, Alizadeh, Kraska, Papaemmanouil, and Tatbul}{Marcus et~al\mbox{.}}{2019}]%
        {marcus2019neo}
\bibfield{author}{\bibinfo{person}{Ryan Marcus}, \bibinfo{person}{Parimarjan Negi}, \bibinfo{person}{Hongzi Mao}, \bibinfo{person}{Chi Zhang}, \bibinfo{person}{Mohammad Alizadeh}, \bibinfo{person}{Tim Kraska}, \bibinfo{person}{Olga Papaemmanouil}, {and} \bibinfo{person}{Nesime Tatbul}.} \bibinfo{year}{2019}\natexlab{}.
\newblock \showarticletitle{Neo: A learned query optimizer}.
\newblock \bibinfo{journal}{\emph{arXiv preprint arXiv:1904.03711}} (\bibinfo{year}{2019}).
\newblock


\bibitem[\protect\citeauthoryear{Mou, Li, Zhang, Wang, and Jin}{Mou et~al\mbox{.}}{2016}]%
        {mou2016convolutional}
\bibfield{author}{\bibinfo{person}{Lili Mou}, \bibinfo{person}{Ge Li}, \bibinfo{person}{Lu Zhang}, \bibinfo{person}{Tao Wang}, {and} \bibinfo{person}{Zhi Jin}.} \bibinfo{year}{2016}\natexlab{}.
\newblock \showarticletitle{Convolutional neural networks over tree structures for programming language processing}. In \bibinfo{booktitle}{\emph{Proceedings of the AAAI conference on artificial intelligence}}, Vol.~\bibinfo{volume}{30}.
\newblock


\bibitem[\protect\citeauthoryear{Naveed, Khan, Qiu, Saqib, Anwar, Usman, Akhtar, Barnes, and Mian}{Naveed et~al\mbox{.}}{2023}]%
        {naveed2023comprehensive}
\bibfield{author}{\bibinfo{person}{Humza Naveed}, \bibinfo{person}{Asad~Ullah Khan}, \bibinfo{person}{Shi Qiu}, \bibinfo{person}{Muhammad Saqib}, \bibinfo{person}{Saeed Anwar}, \bibinfo{person}{Muhammad Usman}, \bibinfo{person}{Naveed Akhtar}, \bibinfo{person}{Nick Barnes}, {and} \bibinfo{person}{Ajmal Mian}.} \bibinfo{year}{2023}\natexlab{}.
\newblock \showarticletitle{A comprehensive overview of large language models}.
\newblock \bibinfo{journal}{\emph{arXiv preprint arXiv:2307.06435}} (\bibinfo{year}{2023}).
\newblock


\bibitem[\protect\citeauthoryear{Negi, Wu, Kipf, Tatbul, Marcus, Madden, Kraska, and Alizadeh}{Negi et~al\mbox{.}}{2023}]%
        {negi2023robust}
\bibfield{author}{\bibinfo{person}{Parimarjan Negi}, \bibinfo{person}{Ziniu Wu}, \bibinfo{person}{Andreas Kipf}, \bibinfo{person}{Nesime Tatbul}, \bibinfo{person}{Ryan Marcus}, \bibinfo{person}{Sam Madden}, \bibinfo{person}{Tim Kraska}, {and} \bibinfo{person}{Mohammad Alizadeh}.} \bibinfo{year}{2023}\natexlab{}.
\newblock \showarticletitle{Robust query driven cardinality estimation under changing workloads}.
\newblock \bibinfo{journal}{\emph{Proceedings of the VLDB Endowment}} \bibinfo{volume}{16}, \bibinfo{number}{6} (\bibinfo{year}{2023}), \bibinfo{pages}{1520--1533}.
\newblock


\bibitem[\protect\citeauthoryear{OpenAI}{OpenAI}{2023}]%
        {openai2023gpt4}
\bibfield{author}{\bibinfo{person}{OpenAI}.} \bibinfo{year}{2023}\natexlab{}.
\newblock \bibinfo{title}{GPT-4 Technical Report}.
\newblock
\newblock
\showeprint[arxiv]{2303.08774}~[cs.CL]


\bibitem[\protect\citeauthoryear{Ouyang, Wu, Jiang, Almeida, Wainwright, Mishkin, Zhang, Agarwal, Slama, Ray, et~al\mbox{.}}{Ouyang et~al\mbox{.}}{2022}]%
        {ouyang2022training}
\bibfield{author}{\bibinfo{person}{Long Ouyang}, \bibinfo{person}{Jeffrey Wu}, \bibinfo{person}{Xu Jiang}, \bibinfo{person}{Diogo Almeida}, \bibinfo{person}{Carroll Wainwright}, \bibinfo{person}{Pamela Mishkin}, \bibinfo{person}{Chong Zhang}, \bibinfo{person}{Sandhini Agarwal}, \bibinfo{person}{Katarina Slama}, \bibinfo{person}{Alex Ray}, {et~al\mbox{.}}} \bibinfo{year}{2022}\natexlab{}.
\newblock \showarticletitle{Training language models to follow instructions with human feedback}.
\newblock \bibinfo{journal}{\emph{Advances in neural information processing systems}}  \bibinfo{volume}{35} (\bibinfo{year}{2022}), \bibinfo{pages}{27730--27744}.
\newblock


\bibitem[\protect\citeauthoryear{Pourreza and Rafiei}{Pourreza and Rafiei}{2024}]%
        {pourreza2024din}
\bibfield{author}{\bibinfo{person}{Mohammadreza Pourreza} {and} \bibinfo{person}{Davood Rafiei}.} \bibinfo{year}{2024}\natexlab{}.
\newblock \showarticletitle{Din-sql: Decomposed in-context learning of text-to-sql with self-correction}.
\newblock \bibinfo{journal}{\emph{Advances in Neural Information Processing Systems}}  \bibinfo{volume}{36} (\bibinfo{year}{2024}).
\newblock


\bibitem[\protect\citeauthoryear{Shi, Tang, Zhang, Zhang, and Yang}{Shi et~al\mbox{.}}{2024}]%
        {shi2024survey}
\bibfield{author}{\bibinfo{person}{Liang Shi}, \bibinfo{person}{Zhengju Tang}, \bibinfo{person}{Nan Zhang}, \bibinfo{person}{Xiaotong Zhang}, {and} \bibinfo{person}{Zhi Yang}.} \bibinfo{year}{2024}\natexlab{}.
\newblock \showarticletitle{A survey on employing large language models for text-to-sql tasks}.
\newblock \bibinfo{journal}{\emph{arXiv preprint arXiv:2407.15186}} (\bibinfo{year}{2024}).
\newblock


\bibitem[\protect\citeauthoryear{Stanley}{Stanley}{2015}]%
        {stanley2015catalan}
\bibfield{author}{\bibinfo{person}{Richard~P Stanley}.} \bibinfo{year}{2015}\natexlab{}.
\newblock \bibinfo{booktitle}{\emph{Catalan numbers}}.
\newblock \bibinfo{publisher}{Cambridge University Press}.
\newblock


\bibitem[\protect\citeauthoryear{Touvron, Martin, Stone, Albert, Almahairi, Babaei, Bashlykov, Batra, Bhargava, Bhosale, et~al\mbox{.}}{Touvron et~al\mbox{.}}{2023}]%
        {touvron2023llama}
\bibfield{author}{\bibinfo{person}{Hugo Touvron}, \bibinfo{person}{Louis Martin}, \bibinfo{person}{Kevin Stone}, \bibinfo{person}{Peter Albert}, \bibinfo{person}{Amjad Almahairi}, \bibinfo{person}{Yasmine Babaei}, \bibinfo{person}{Nikolay Bashlykov}, \bibinfo{person}{Soumya Batra}, \bibinfo{person}{Prajjwal Bhargava}, \bibinfo{person}{Shruti Bhosale}, {et~al\mbox{.}}} \bibinfo{year}{2023}\natexlab{}.
\newblock \showarticletitle{Llama 2: Open foundation and fine-tuned chat models}.
\newblock \bibinfo{journal}{\emph{arXiv preprint arXiv:2307.09288}} (\bibinfo{year}{2023}).
\newblock


\bibitem[\protect\citeauthoryear{Vaswani}{Vaswani}{2017}]%
        {vaswani2017attention}
\bibfield{author}{\bibinfo{person}{A Vaswani}.} \bibinfo{year}{2017}\natexlab{}.
\newblock \showarticletitle{Attention is all you need}.
\newblock \bibinfo{journal}{\emph{Advances in Neural Information Processing Systems}} (\bibinfo{year}{2017}).
\newblock


\bibitem[\protect\citeauthoryear{Xu, Alon, Neubig, and Hellendoorn}{Xu et~al\mbox{.}}{2022}]%
        {xu2022systematic}
\bibfield{author}{\bibinfo{person}{Frank~F Xu}, \bibinfo{person}{Uri Alon}, \bibinfo{person}{Graham Neubig}, {and} \bibinfo{person}{Vincent~Josua Hellendoorn}.} \bibinfo{year}{2022}\natexlab{}.
\newblock \showarticletitle{A systematic evaluation of large language models of code}. In \bibinfo{booktitle}{\emph{Proceedings of the 6th ACM SIGPLAN International Symposium on Machine Programming}}. \bibinfo{pages}{1--10}.
\newblock


\bibitem[\protect\citeauthoryear{Xu, Ding, Hu, Niemier, Cong, Hu, and Shi}{Xu et~al\mbox{.}}{2018}]%
        {xu2018scaling}
\bibfield{author}{\bibinfo{person}{Xiaowei Xu}, \bibinfo{person}{Yukun Ding}, \bibinfo{person}{Sharon~Xiaobo Hu}, \bibinfo{person}{Michael Niemier}, \bibinfo{person}{Jason Cong}, \bibinfo{person}{Yu Hu}, {and} \bibinfo{person}{Yiyu Shi}.} \bibinfo{year}{2018}\natexlab{}.
\newblock \showarticletitle{Scaling for edge inference of deep neural networks}.
\newblock \bibinfo{journal}{\emph{Nature Electronics}} \bibinfo{volume}{1}, \bibinfo{number}{4} (\bibinfo{year}{2018}), \bibinfo{pages}{216--222}.
\newblock


\bibitem[\protect\citeauthoryear{Yang, Chiang, Luan, Mittal, Luo, and Stoica}{Yang et~al\mbox{.}}{2022}]%
        {yang2022balsa}
\bibfield{author}{\bibinfo{person}{Zongheng Yang}, \bibinfo{person}{Wei-Lin Chiang}, \bibinfo{person}{Sifei Luan}, \bibinfo{person}{Gautam Mittal}, \bibinfo{person}{Michael Luo}, {and} \bibinfo{person}{Ion Stoica}.} \bibinfo{year}{2022}\natexlab{}.
\newblock \showarticletitle{Balsa: Learning a query optimizer without expert demonstrations}. In \bibinfo{booktitle}{\emph{Proceedings of the 2022 International Conference on Management of Data}}. \bibinfo{pages}{931--944}.
\newblock


\bibitem[\protect\citeauthoryear{Yu, Chai, Li, and Liu}{Yu et~al\mbox{.}}{2022}]%
        {yu2022cost}
\bibfield{author}{\bibinfo{person}{Xiang Yu}, \bibinfo{person}{Chengliang Chai}, \bibinfo{person}{Guoliang Li}, {and} \bibinfo{person}{Jiabin Liu}.} \bibinfo{year}{2022}\natexlab{}.
\newblock \showarticletitle{Cost-based or learning-based? A hybrid query optimizer for query plan selection}.
\newblock \bibinfo{journal}{\emph{Proceedings of the VLDB Endowment}} \bibinfo{volume}{15}, \bibinfo{number}{13} (\bibinfo{year}{2022}), \bibinfo{pages}{3924--3936}.
\newblock


\bibitem[\protect\citeauthoryear{Zhao, Cong, Shi, and Miao}{Zhao et~al\mbox{.}}{2022}]%
        {zhao2022queryformer}
\bibfield{author}{\bibinfo{person}{Yue Zhao}, \bibinfo{person}{Gao Cong}, \bibinfo{person}{Jiachen Shi}, {and} \bibinfo{person}{Chunyan Miao}.} \bibinfo{year}{2022}\natexlab{}.
\newblock \showarticletitle{Queryformer: A tree transformer model for query plan representation}.
\newblock \bibinfo{journal}{\emph{Proceedings of the VLDB Endowment}} \bibinfo{volume}{15}, \bibinfo{number}{8} (\bibinfo{year}{2022}), \bibinfo{pages}{1658--1670}.
\newblock


\bibitem[\protect\citeauthoryear{Zhu, Chen, Ding, Chen, Pfadler, Wu, and Zhou}{Zhu et~al\mbox{.}}{2023}]%
        {zhu2023lero}
\bibfield{author}{\bibinfo{person}{Rong Zhu}, \bibinfo{person}{Wei Chen}, \bibinfo{person}{Bolin Ding}, \bibinfo{person}{Xingguang Chen}, \bibinfo{person}{Andreas Pfadler}, \bibinfo{person}{Ziniu Wu}, {and} \bibinfo{person}{Jingren Zhou}.} \bibinfo{year}{2023}\natexlab{}.
\newblock \showarticletitle{Lero: A learning-to-rank query optimizer}.
\newblock \bibinfo{journal}{\emph{Proceedings of the VLDB Endowment}} \bibinfo{volume}{16}, \bibinfo{number}{6} (\bibinfo{year}{2023}), \bibinfo{pages}{1466--1479}.
\newblock


\end{thebibliography}

\end{document}